\documentclass[aps,pra,twocolumn,
					showpacs,superscriptaddress,nofootinbib]{revtex4-1}

\usepackage{helvet}
\usepackage{eurosym}
\usepackage{color}
\usepackage[latin1]{inputenc}
\usepackage{amssymb}
\usepackage{amsmath}
\usepackage{graphicx}
\usepackage{bm}
\usepackage{xcolor}

\usepackage{natbib}
\usepackage{url}
\usepackage{textcase}
\usepackage{bm}

\DeclareGraphicsExtensions{%
 .png,%
 .jpg,%
 .pdf,.PDF,%
 .mps,.jpeg,.jbig2,.jb2,.JPG,.JPEG,.JBIG2,.JB2}

\newcommand{\abs}[1]{\left| #1 \right|} 
\newcommand{\avg}[1]{\left< #1 \right>} 

\newcommand{\ket}[1]{\left| #1 \right>} 
\newcommand{\bra}[1]{\left< #1 \right|} 
\newcommand{\braket}[2]{\left< #1 \vphantom{#2} \right|
 \left. #2 \vphantom{#1} \right>} 
\newcommand{\rb}[1]{\left( #1 \right)}
\newcommand{\rbb}[1]{\left[ #1 \right]}

\newcommand{\commut}[1]{\left[ #1 \right]}

\begin{document}
\title{Nonadiabatic dynamics of the excited states for the Lipkin-Meshkov-Glick model}
\author{Wassilij Kopylov}
\affiliation{Institut f\"ur Theoretische Physik,
 Technische Universit\"at Berlin,
 D-10623 Berlin,
 Germany}
\author{Gernot Schaller}
\affiliation{Institut f\"ur Theoretische Physik,
  Technische Universit\"at Berlin,
  D-10623 Berlin,
  Germany}
\author{Tobias Brandes}
\affiliation{Institut f\"ur Theoretische Physik,
 Technische Universit\"at Berlin,
 D-10623 Berlin,
 Germany}
\date{\today}

\begin{abstract}
We theoretically investigate the impact of the excited state quantum phase transition  on the adiabatic dynamics for the Lipkin-Meshkov-Glick model. Using a time dependent protocol, we continuously change a model parameter and then discuss the scaling properties of the system especially close to the excited state quantum phase transition where we find that these depend on the energy eigenstate. On top, we show that the mean-field dynamics with the time dependent protocol gives the correct scaling and expectation values in the thermodynamic limit even for the excited states. 
\end{abstract}

\maketitle


\section{Introduction}
Universality and feature analogy connect different branches of physics. One of the most famous examples are phase transitions which span from cosmology to the quantum world \cite{Phase_Transition-Universal_dynamics_topologica_defects-Zurek,Cosmology_Experiments_in_helium-Zurek,Phase_Transition-Theory_of_critical_phenomena-Hohenberg}. Here one of the intriguing questions are scaling properties around the phase transition in both equilibrium and non-equilibrium setups. One way to create the latter is a continuous quench from one state to another \cite{Review-Noneq_dynamics_closed_systems-Vengalattore,QuantSys-Equilibration_thermalization-Yukalov}. The connection between both can be characterized by the Kibble-Zurek mechanism,  which connects the quench-induced defect formation with the universal scaling exponents in equilibrium such as relaxation time and the healing length \cite{Cosmology-phase_transition-Kibble,Cosmology-Experiments_in_condensed_matter-Zurek}.

A quantum phase transition (QPT) is a paradigm at the small-scale domain, its famous property is the non-analyticity of the ground-state energy and ground-state wave-function at some point in the system configuration space. 
The validity of the Kibble-Zurek mechanism has been applied to QPTs in the Ising-like models with a finite dimension \cite{QPT-Dynamics-Zoller,Ising-Dynamics_of_QPT_exact_solution-Dziarmaga,QPT-Networks_addiabatic-Johnson}. Later it was shown that the scaling connection still holds even in zero-dimensional models like the Dicke \cite{Dicke-Dicke_Modell}, where the length-scale is absent \cite{Rabi-QPT-addiab_dynam-plenio,Dicke-Robust_quantum_correlation_with_linear_increased_coupling-Acevedo}. Furthermore, its connection to Landau-Zener transitions has been shown \cite{Landau-Zur_theory_der_energieuebertragung,Zener-non_adiabatic_crossing_of_energy_levels,Phase_Transition-Universal_dynamics_topologica_defects-Zurek,Kibble-Zurek-siplest_model_and_landau_zener_connection-Bogdan}. 
The scaling connection is relevant e.g. for quantum computation algorithms in systems with a QPT \cite{quantum_comput-Decoherence_in_qpt-Gernot,quantum_comput-decoherence_in_dynamic_qpt_ising-Gernot}. Note, that the Kibble-Zurek mechanism has been verified in various experimental setups e.g. using quantum atomic gases  \cite{Kibble-Zurek-Exp_Dark_soliton_and_matter_wave-Oberthaler,Kibble-Zureck-Exp_quench_in_bose_gas-Beugnon,Kibble_Zureck-Experiment_coherence_dynamics_qpt_Bloch,Kibble-Zureck-inverse_scenario_with_bose_gases-Yukalov}, superfluids \cite{Kibble_Zurek-Exp_somic_string_laboratary_simulation-Pickett}  or Coulomb Crystals \cite{Kibble-Zureck-Exp_Coulomb_Crystals-Reznik}. 

Here we extend the study of scaling properties toward excited state quantum phase transition (ESQPT) \cite{ESQPT-inmany_body_systems-Caprio,ESQPT_in_quantum_optical_models-Pedro,brandes2013excited,ESQPT-flow-level_dynamics-thermal_properties-cejnar,ESQPT-system_with_two_freedom_degrees_finite_size-Cejnar}. Similar to the ground state QPT, the ESQPT is defined by a discontinuity in the density of states in the excited part of the spectrum, it is imprinted in different observables and was experimentally found e.g. in the bending spectra of different molecular systems \cite{ESQPT-exp_molecular_system-Koput,ESQPT-exp_molecular_system-Iachello,ESQPT-exp_vibration_bending_dynamics_triatomic_molecules-Iachello}. 

For our investigation we consider the so-called Lipkin-Meshkov-Glick (LMG) model, which describes the dynamics of $N$ two-level systems with an internal all-to-all coupling. This model was first introduced in the context of nuclear physics to check the validity of different predictions \cite{LMG-lipkin1965validity,LMG-validity_many_body_approx-Lipkin,LMG-validity_many_body_approx-Lipkin-3}. Recently, it has attracted much attention again in the context of cold atoms and many-body physics, especially as a rather simple model with an experimental realization for studying QPTs \cite{LMG-Exp_Bifurcation_rabi_to_jesophson-Oberthaler,LMG_Exp_nonlinear_atom_interferometer-Oberthaler,LMG-Analytical_solution-Feng,LMG-spectrum_thermodynamic_limit_and_finite_size-corr-Mosseri,LMG-Exp_non-local_propagation_of_corr-Monroe}.  

The QPT-induced scaling properties, the excited state impact to the wave-package dynamics or the transitionless driving in the time-dependent LMG model in presence of a linear or a sudden quench has been theoretically investigated in Ref.  \cite{LMG-adiabatic_dynamics_Caneva,LMG-dynamical_properties_accros_qpt-Mosseri,ESQPT-quantum-quench-pedro,LMG-Shortcut_to_adiabaticity-Campbell,LMG-irreversible_processes_without_energy_dissipation-Relano,ESQPT-Structure_eigenstate_and_quench_dynamic-Santos,ESQPT-manybody_localization_dynamics_bifurcation-Santos}, but  with the ground state as the initial condition. The ground state is a common choice, especially in the context of the quantum computation and annealing problems \cite{Cavity-quantum_annealing_network_kerr_oszi-Grimsmo,quantum_comput-general_error-Gernot}. In contrast, here we choose several excited states as the initial condition to probe the scaling effects induced by the ESQPT and find that the scaling parameter decreases and converges to a constant value if the energy of the chosen initial excited state increases. Note, that our study has experimental relevance, as the excited states can be prepared with coherent states which are commonly used in the existing experimental setup of the LMG model \cite{LMG-Exp_Bifurcation_rabi_to_jesophson-Oberthaler,LMG_Exp_nonlinear_atom_interferometer-Oberthaler}. 

Our work is organized as follows. In Sec. \ref{sec:1_model} we introduce the model, in Sec. \ref{sec:1-2-spectral} we review its spectral characteristics and investigate the gap closing scaling for the excited states. In Sec. \ref{sec:2-add-dynam}  we numerically investigate the scaling properties due to the ESQPT and in Sec. \ref{sec:3-state-dynamics} the state dynamics for different speeds of the quench protocol by solving the time-dependent Schr\"{o}dinger equation. In Sec. \ref{sec:meanfield} we show that the results from the previous chapters can be obtained form the corresponding mean-field model. We summarize our findings in Sec.\ref{sec:5-summary}.

\section{Model}
\label{sec:1_model}
The anisotropic LMG Hamiltonian reads \cite{LMG-Critical_scaling_law_entaglement-Vidal}

\begin{equation}
\label{eq:lmg_ham}
\hat{H} = -h \hat{J}_z - \frac{\gamma_x}{N} \hat{J}_x^2,
\end{equation}

where $\hat{J}_\eta = \frac{1}{2} \sum_{i=1}^N \hat{\sigma}_\eta^{(i)}$, $\eta = \{x,y,z\}$, are collective angular momentum operators and $\hat{\sigma}_{x,y,z}$ are the common Pauli matrices. The magnetic field $h$ defines the level splitting in the non-interacting case, whereas $\gamma_x$ describes the coupling strength between the $N$ two-level systems. Note, that here the angular momentum $j$ is directly connected to the system size via $N = 2j$, this suggests to use the angular momentum basis of the $\hat{J_z}$ operator $\ket{j,m}$ for the matrix representation of the Hamiltonian Eq. \eqref{eq:lmg_ham}.

The Hamiltonian Eq. \eqref{eq:lmg_ham} preserves the length of the total spin $\hat{J}^2 = \hat{J}_x^2 + \hat{J}_y^2 + \hat{J}_z^2$ \cite{LMG-large_N_scaling_behaviour-Heiss} and has a parity symmetry which divides $\hat{H}$ into two different independent energetic subspaces, one even '+' and one odd '--' \cite{LMG-from_perspective_of_SU11-model-Dukelsky,LMG-networks_qpt-sorokin}. The corresponding parity operator reads 

\begin{equation}
\label{eq:lmg_parity}
\hat{P}_\pm =  \frac{1}{2} \rbb{ \mathbf{1} \pm \exp\rb{i \pi \hat{J}_z} \exp\rb{i \pi N/2}}.
\end{equation}

Note, the parity conservation is directly imprinted in the matrix representation of the Hamiltonian Eq. \eqref{eq:lmg_ham} 

\begin{align}
&\bra{j,m'}\hat{H}\ket{j,m} \\
&= \rbb{-h m - \frac{\gamma_x}{2 N}\rb{k_-(j,m)^2+k_+(j,m)^2} }\delta_{m',m} \notag \\
& \quad - \frac{\gamma_x}{2 N}k_+(j,m)k_+(j,m+1) \delta_{m',m+2}  \notag \\
& \quad -\frac{\gamma_x}{2 N} k_-(j,m)k_-(j,m-1) \delta_{m',m-2} \notag 
\end{align}
as a decomposition into two decoupled regions with even or odd state $\ket{j,m}$. Here $\hat{J}_\pm\ket{j,m}=k_\pm(j,m)\ket{j,m \pm 1}$ with $\hat{J}_\pm = \hat{J}_x \pm i \hat{J}_y$ is used. 

As parity is conserved, we will restrict ourselves only to the even '+' parity and will perform all calculations with the even Hamiltonian  

\begin{equation}
\label{eq:lmg_ham_+}
\hat{H}_+ = \hat{P}_+^\dag \hat{H} \hat{P}_+,
\end{equation}
which contains the ground state of the system. In the following, we will denote the eigenstates of the Hamiltonian $\hat{H}_+$ by $\ket{i}$ for fixed $\gamma_x$ values, i.e.
\begin{equation}
\label{eq:H+_eigenvalues}
\hat{H}_+ \ket{i} = E^+_i \ket{i}.
\end{equation}

Note that the spectra of both even and odd Hamiltonians become identical in the $N \to \infty$ limit. The lower part of the even-parity spectrum is shown in Fig. \ref{fig:energy_spectrum} as a function of the coupling strength $\gamma_x$.

\section{Spectral Properties}
\label{sec:1-2-spectral}

\begin{figure}[b]
\includegraphics[width= \columnwidth]{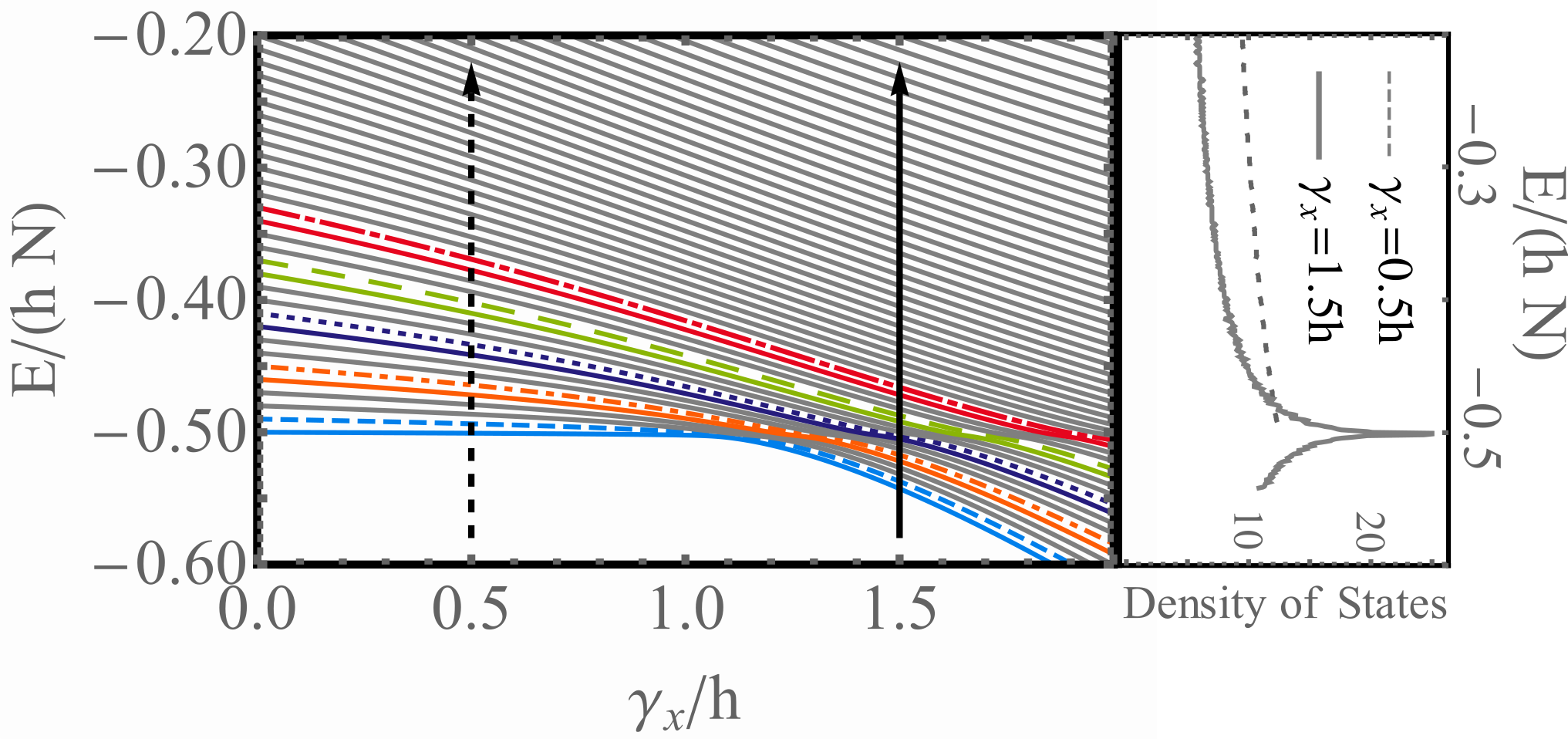}
\caption{(color online)(left panel) The even part of the energy spectrum for the LMG model Eq. \eqref{eq:lmg_ham_+} for a finite atomic number N. A few of the low-lying neighboring states are marked with solid and dashed color lines. Parameters: $N = 200$. (right panel) From the spectrum numerically determined density of states for two different $\gamma_x$ values, along the dashed and solid arrows in the left panel. The logarithmic peak (solid line) appears for $\gamma_x/h >\gamma_x^{cr,1}$ and indicates the ESQPT at $E/(hN) = 1/2$.  Parameters: $N=10000$. }
\label{fig:energy_spectrum}
\end{figure}

\begin{figure}
\includegraphics[width=0.95 \columnwidth]{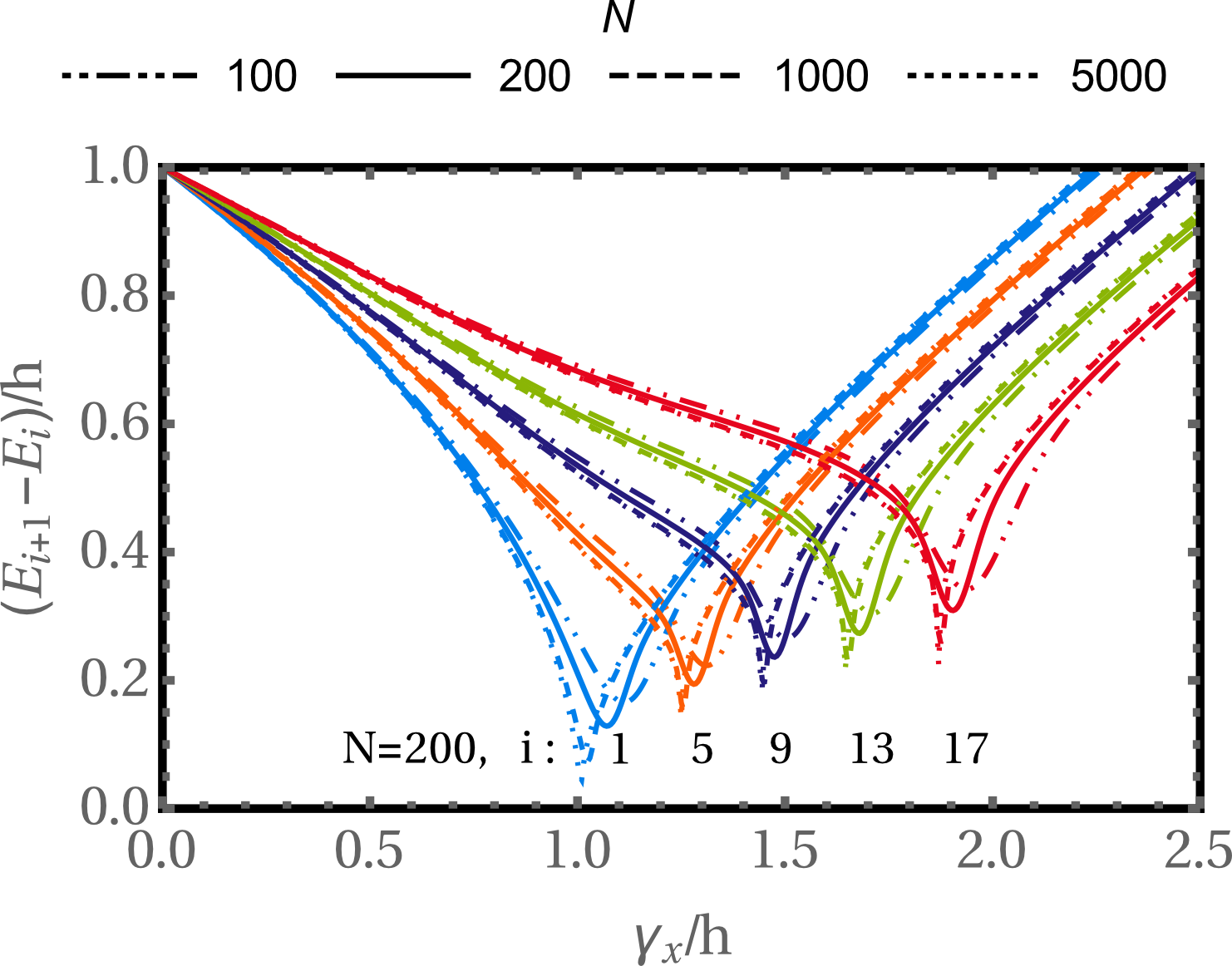}
\includegraphics[width=0.95 \columnwidth]{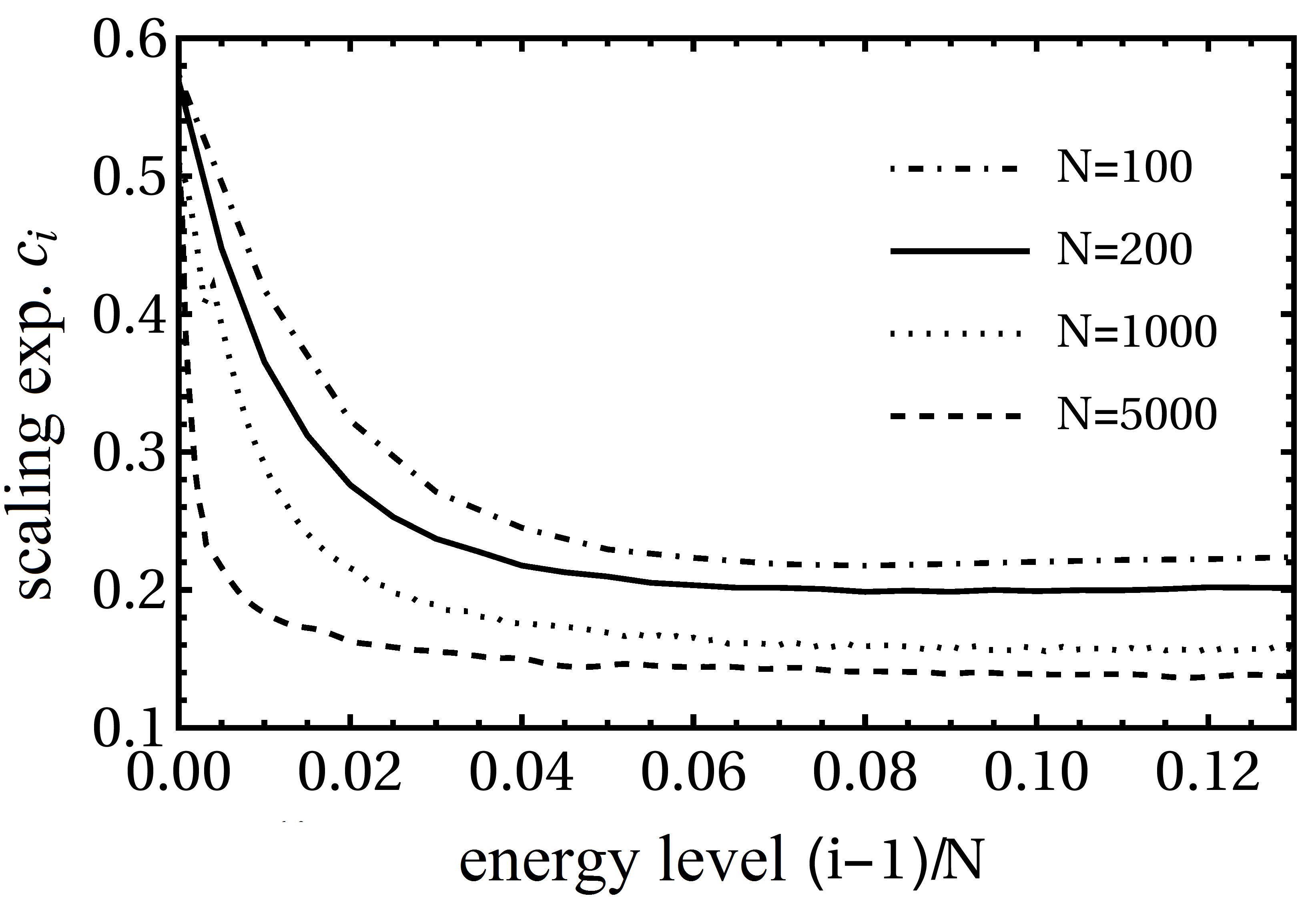}
\caption{(color online)(top panel) The energy difference between neighboring eigenstates $\ket{i+1}$ and $\ket{i}$ of the $\hat{H}_+$ Hamiltonian \eqref{eq:lmg_ham_+} as a function of $\gamma_x/h$ for different numbers $N$. The neighboring states shown here are marked in Fig. \ref{fig:energy_spectrum} with the same color for $N=200$. The shown state  numbers $i$ numbers is for $N = 200$. In case of a different $N$ value, the correspondent state number has to be rescaled according to $(i-1)\cdot N/200 +1$.  The gap closing shows the value for $\gamma_x^{cr,i}/h$ where the ESQPT takes place. (lower panel) The exponent $c_i$, Eq. \eqref{eq:peak_fit}, characterizes the gap closing for a state $\ket{i}$ for different particle numbers $N$. Note, the relative scaling of the x-axis.}
\label{fig:energy_spectrum_gap}
\end{figure}

A lot of intriguing physical properties are incorporated in Eq. \eqref{eq:lmg_ham}. Most prominently, a second order quantum phase transition (QPT) appears in the thermodynamic limit $N \to \infty$ at $\gamma_x/h = \gamma_x^{cr,1}/h \equiv 1$ \cite{LMG-phase_transition-Gilmore,LMG-grouns_state_studies_coherent_states-Gilmore,LMG-spectrum_thermodynamic_limit_and_finite_size-corr-Mosseri}, which is visible in the energy spectrum. There, the ground state energy becomes non-analytic and the gap to the first excited state closes as a function of $\gamma_x/h$, see the two blue (solid and dashed) lines in Fig.~\ref{fig:energy_spectrum} (left panel). The energy distance between the ground and the first excited state is shown in Fig.~\ref{fig:energy_spectrum_gap} (top panel) by blue colored lines for different particle numbers $N$. For $N \to \infty$ the ground state gap closes and the position of the minimum of the $i=1$ curves converges to $\gamma_x^{cr,1}/h = 1$. Note that the energy axis in Fig. \ref{fig:energy_spectrum_gap} is not scaled with $1/N$ , unlike the energy axis in Fig. \ref{fig:energy_spectrum}. The energy spectrum was determined by a numerical diagonalization of $\hat{H}_+$ in Eq. \eqref{eq:lmg_ham_+} represented in the $\hat{J}_z$ basis $\{\ket{m}\}$, i.e.
\begin{equation}
\hat{J}_z \ket{m} = m \ket{m}.
\end{equation}

In the thermodynamic limit the excited spectrum possesses  an additional phase transition - the excited state quantum phase transition (ESQPT) \cite{LMG-large_N_scaling_behaviour-Heiss,ESQPT-inmany_body_systems-Caprio,ESQPT_Docoherence_two_level_boson_pedro,Bastidas-Quantum_criticality_in_kicked_top,LMG-TC-periodic_dynamic_and_QPT-Georg}. In the case of the LMG model, the ESQPT is visible as a logarithmic divergence of the spectral level density at the energy $E = -0.5/(h \cdot N)$ of the so called separatrix, which existence is typical for a ESQPT \cite{ESQPT_imparct_of_qpt_on_level_dynamics-Cejnar,QPT_interacting_boson_model-Cejnar}, for a fixed $\gamma_x > \gamma_x^{cr,1}$ value \cite{LMG-thermodynamical_limit-Mosseri}. In the right panel of Fig. \ref{fig:energy_spectrum} we plot the numerically determined level density for two different $\gamma_x$ values. Whereas for $\gamma_x < \gamma_x^{cr,1}$ the density of states is approximately flat (dashed curve), for $\gamma_x > \gamma_x^{cr,1}$ the density divergence is clearly visible (solid curve) at $E = -0.5/(h \cdot N)$ A look into the level clustering for finite particle number $N$ visualizes the formation of the ESQPT, see Fig.~\ref{fig:energy_spectrum}. It is build by an infinite series of avoided level crossings of the neighboring excited states, as their gap closes at the energy of the ESQPT in the thermodynamic limit \cite{LMG-dynamical_properties_accros_qpt-Mosseri,ESQPT-inmany_body_systems-Caprio,ESQPT-system_with_two_freedom_degrees_finite_size-Cejnar,ESQPT-from_nonhermitian_perspective-Moiseyev,LMG-from_perspective_of_SU11_esqpt-dukelsky}. Some of such neighboring states are marked by solid and dashed colored lines in the Fig.~\ref{fig:energy_spectrum}.  

We visualize this gap closing by plotting the energy distance between the neighboring excited states $\ket{i}$ and $\ket{i+1}$ which are colored in Fig.~\ref{fig:energy_spectrum_gap} (top panel) for different values of $N$. Qualitatively, the form of the curves for the excited states and their finite-size scaling behavior is similar to the ground state curves for $\ket{i=1}$. The cusp-like minimum of the curve $i$ is at a position $\gamma_x^{cr,i}/h$ which increases with $i$. It marks the region where level density around the state $\ket{i}$ diverges in the thermodynamic limit and the ESQPT takes place. In the $E(\gamma_x)$ representation, the flow of the states $\ket{i}$ and $\ket{i+1}$ will bring them  to the energy of $E(\gamma_x^{cr,i})=-0.5/N$ at $\gamma_x^{cr,i}/h$, see Fig.~\ref{fig:energy_spectrum} \cite{ESQPT-flow-level_dynamics-thermal_properties-cejnar}. For a further analysis of the gap closing in the excited states we will later-on define an adiabatic protocol approaching the point $\gamma_x^{cr,i}$ from the left. Therefore, we fit the left part of each curve $i$ in Fig.~\ref{fig:energy_spectrum_gap} (top panel)  close to the cusp minimum  by \cite{Sachdev-QPT} 
\begin{equation}
\label{eq:peak_fit}
E_{i+1} - E_i = a_i \cdot \abs{\gamma_x-\gamma_x^{cr,i}}^{-c_i}
\end{equation}
for different $N$ values, here the $a_i,c_i$ are the fitting parameters and $\gamma_x^{cr,i}$ is determined numerically by finding the energetic minimum of the difference $E_{i+1}-E_{i}$ by varying $\gamma_x$. The obtained numbers for the critical scaling $c_i$ of the states $\ket{i+1}$ and $\ket{i}$ are shown in Fig.~\ref{fig:energy_spectrum_gap}(lower panel). The scaling behavior is quantitatively similar for different particle numbers $N$. The ground state scaling parameter $c_{i=1}$ approaches $1/2$ for $N \to \infty$, which is typical for a second order QPT \cite{Sachdev-QPT}. For the excited states $\ket{i>1}$ $c_i$ decays till a constant value. Hence, the energy distance of higher neighboring states $i$ and $i+1$ develops close to the ESQPT region in a similar manner.

\section{Adiabatic Dynamics}
\label{sec:2-add-dynam}
We now  analyse the system behavior by continuously quenching it through the ESQPT, starting in an excited state of $\hat{H}_+$. To probe the influence of the ESQPT on the adiabatic system dynamics, we assume a time dependent interaction strength $\gamma_x(t)$, linearly increased from some starting value $\gamma^{(i)}$ at time $t_{\rm in}$ to the final value $\gamma^{(f)}$ at time $t_f$

\begin{equation}
\label{eq:gamma-timedep}
\gamma_x(t) = \frac{t}{Q},
\end{equation}
where the quench rate $Q$ determines the inverse speed of the protocol, and $t \in [t_{\rm in},t_f]$. Note that the ground state QPT for our model, Eq. \eqref{eq:lmg_ham}, has been investigated with  similar schemes \cite{LMG-adiabatic_dynamics_Caneva,LMG-Shortcut_to_adiabaticity-Campbell,LMG-dynamical_properties_accros_qpt-Mosseri}.

\begin{figure}
\includegraphics[width=1 \columnwidth]{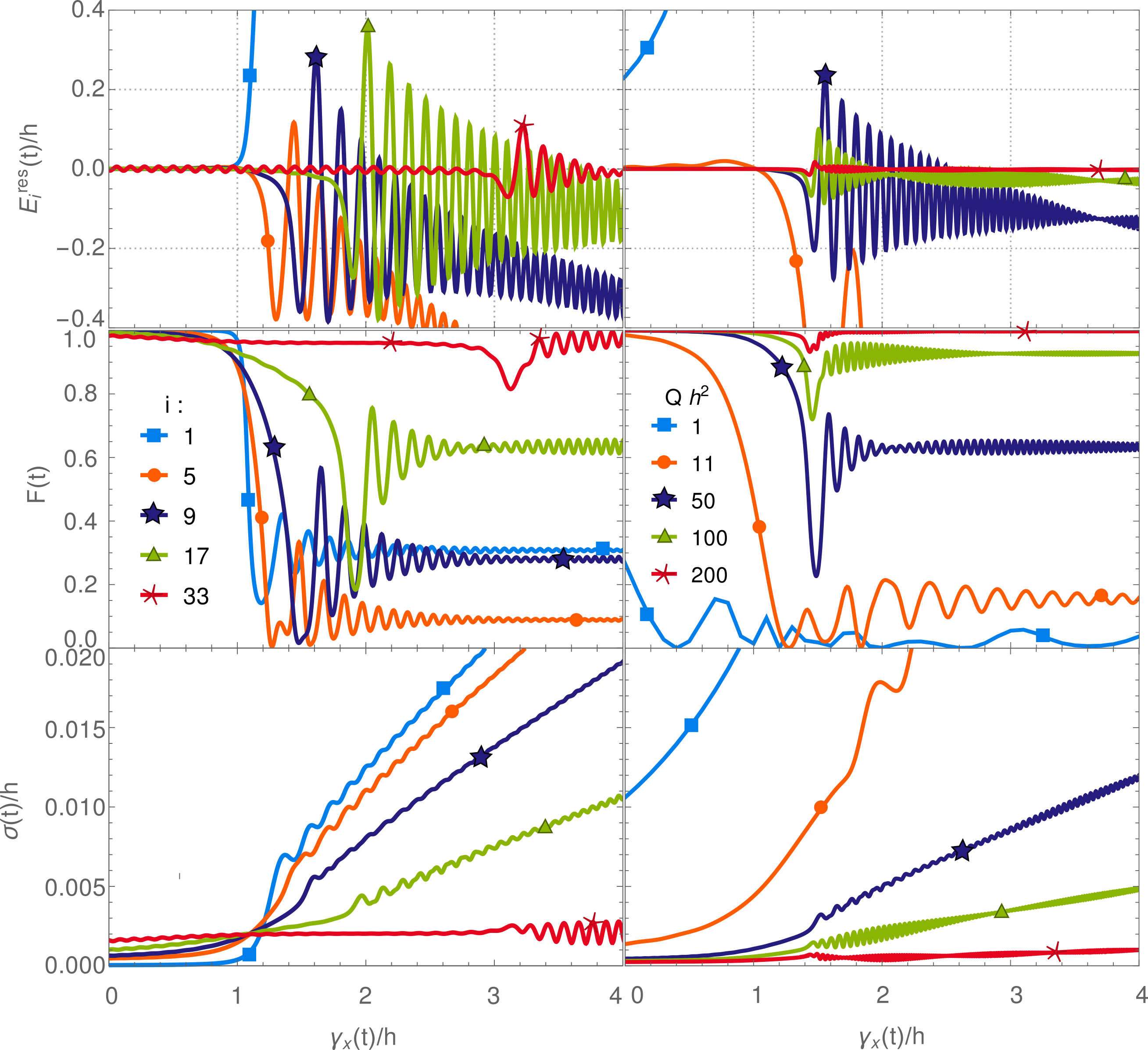}
\caption{(color online) System properties under a continuous linear change of the coupling $\gamma_x$ in time $t$ with a rate $Q$. (top panel) The residual energy $E_i^{res}$, (middle) the fidelity and (lower panel) the distribution width $\sigma$ for different initial eigenstates $\ket{i(t_{\rm in})}$ (left) and different scales $Q$ (right) shows the breakdown of adiabaticity which is connected to the protocol caused transition of the initial wave-package $\ket{i(t_{\rm in})}$ through the ESQPT. Parameters: N = 200, $t_{\rm in} = -10/h$, (left) $Q h^2 = 30$, (right) $\ket{i(t_{\rm in})} = \ket{9}$. }
\label{fig:Lmg_addiab_dyn-N200}
\end{figure}

Thus, starting in a time-local eigenstate $\ket{i(t_{\rm in})}$ of the Hamiltonian $\hat{H}_+(t)$ Eq. \eqref{eq:H+_eigenvalues} at time $t=t_{\rm in}$ with $\gamma = \gamma_x(t_{\rm in})$, we solve the time dependent Schr\"{o}dinger equation 
\begin{equation}
\label{eq:sg}
i \dot{\ket{\psi(t)}} = \hat{H}_+(t) \ket{\psi(t)}, \quad \ket{\psi(t=t_{\rm in})} = \ket{i(t_{\rm in})}.
\end{equation} 

Due to the time dependent Hamiltonian $\hat{H}_+(t)$ the wave package $\Psi(t')$ at time $t' > t_{\rm in}$ will not necessarily be the $i$th eigenstate $\ket{i(t')}$ of the Hamiltonian any more, it will strongly differ especially in the case of the non-adiabatic evolution. We solve this equation numerically by representing the operators in the ${\ket{j,m}}$-Basis.   

To analyze the wave package $\ket{\psi(t)}$, we first introduce the so-called residual energy \cite{LMG-adiabatic_dynamics_Caneva,Rabi-QPT-addiab_dynam-plenio} 
\begin{equation}
\label{eq:def_res_energy}
 E_i^{res}(t) \equiv \bra{\psi(t)}\hat{H}(t)\ket{\psi(t)} - \bra{i(t)}\hat{H}(t)\ket{i(t)},
\end{equation}
which is the difference between the energy expectation value at time $t$ and the adiabatic eigenenergy of the Hamiltonian $\hat{H}_+(t)$ at time $t$. Obviously, this is zero at $t_{\rm in}$ but changes if the adiabatic condition is not fulfilled. If the initial state is the ground state, the residual energy remains positive, but in case of excited states it can have both signs which can switch in time. The top panels of Fig. \ref{fig:Lmg_addiab_dyn-N200} show the residual energy for different initial eigenstates of the Hamiltonian $\hat{H}_+(t)$ at $t_{\rm in}$  and a fixed $Q$ value (left) or for different $Q$ values but a fixed initial eigenstate (right). Note, that we usually choose $t_{\rm in} = -10/h$, therefore $E_i^{res}$ can deviate from  $0$ at $t = 0$, especially for smaller $Q$ values. For higher $Q$ values the deviation is negligible at the begin of the protocol, therefore we show the dynamics only for $\gamma_x(t)/h > 0$ in Fig. \ref{fig:Lmg_addiab_dyn-N200}. The $E_i^{res}$ shows a strongly oscillating behavior which crosses the zero value for all initial states except the ground state which is labeled by '1'. The beginning of the oscillation phase indicates the breakdown of adiabaticity. This is due to the onset of the ESQPT which causes the gap closing between the neighboring excited states, compare with Fig. \ref{fig:energy_spectrum_gap}. 

The wave package $\ket{\psi(t)}$ can be represented in the time-local eigenbasis at each time $t$ 
\begin{equation}
\ket{\psi(t)} = \sum_i \alpha_i(t) \ket{i(t)},
\end{equation}
where $\abs{\alpha_i(t)}^2$ gives the probability that the state $\ket{i(t)}$ is occupied. Note, that the $\alpha_i(t)$ as the wave function $\ket{\psi(t)}$ depend on the choice for the initial condition $\ket{i(t_{\rm in})}$, see Eq. \eqref{eq:sg}. For a better analysis of the adiabaticity loss we need a positive valued signature instead of the residual energy. The simplest choice is the fidelity \cite{LMG-Shortcut_to_adiabaticity-Campbell,ESQPT-manybody_localization_dynamics_bifurcation-Santos} 
\begin{equation}
\label{eq:def_fidelity}
F(t) = \abs{\braket{\psi(t)}{i(t)}}^2.
\end{equation}
It measures the overlap of the state $\ket{i(t)}$ with the wave package and deviates from unity in the non-adiabatic regime. The onset of  deviation is shifted to higher $\gamma_x(t)$ values for higher initial eigenstates and coincides with the beginning of the oscillating phase in the residual energy, see Fig. \ref{fig:Lmg_addiab_dyn-N200} (middle left panel). The protocol quench scale  $Q$, Eq. \eqref{eq:gamma-timedep}, has a strong effect on the adiabaticity loss, too (Fig. \ref{fig:Lmg_addiab_dyn-N200}, middle right panel). For a slower protocol (larger $Q$) the $\gamma_x$ position for the adiabaticity loss shifts closer to the ESQPT value $\gamma_x^{cr,i}$, whereas for a faster protocol the adiabaticity break down shifts to a smaller $\gamma_x$ values, for an initial state $\ket{i(t_{\rm in})}$.  

However, in the ongoing discussion we use a quantity $\sigma$, which is not only positive but contains information about all eigenstates,
\begin{equation}
\label{eq:def_breite_sigma}
\sigma(t) \equiv \frac{1}{N} \sqrt{\sum_j E_j^{res}(t)^2 \abs{\alpha_j(t)}^2},
\end{equation}       
which describes the energy width of the wave package $\ket{\Psi(t)}$ in the time-local basis $\{\ket{i(t)}\}$.~\footnote{Note: The standard deviation of a distribution ${\varepsilon_j}$ with probability $p_j$ is given by $\sqrt{\sum_{j}(\varepsilon_j - \bar{\varepsilon})^2p_j}$ and $\bar{\varepsilon} = \sum_{j} \varepsilon_j p_j$. 
Here, $\bar{\varepsilon} = \bra{\psi}\hat{H}_+\ket{\psi}$.} Similar as in case of the residual energy, Eq. \eqref{eq:def_res_energy} and the fidelity, \eqref{eq:def_fidelity}, the presence of the ESQPT in the excited states affects the dynamical evolution of the $\sigma(t)$, see Fig. \ref{fig:Lmg_addiab_dyn-N200} (lower panels). Its value increases if the adiabaticity is lost, which depends on the position of the ESQPT $\gamma_x^{cr,i}$ and the protocol quench scale $Q$, as we have discussed above. Summarizing, all quantities plotted in Fig. \ref{fig:Lmg_addiab_dyn-N200} consistently display the adiabaticity change.

\begin{figure}
\includegraphics[width=0.99 \columnwidth]{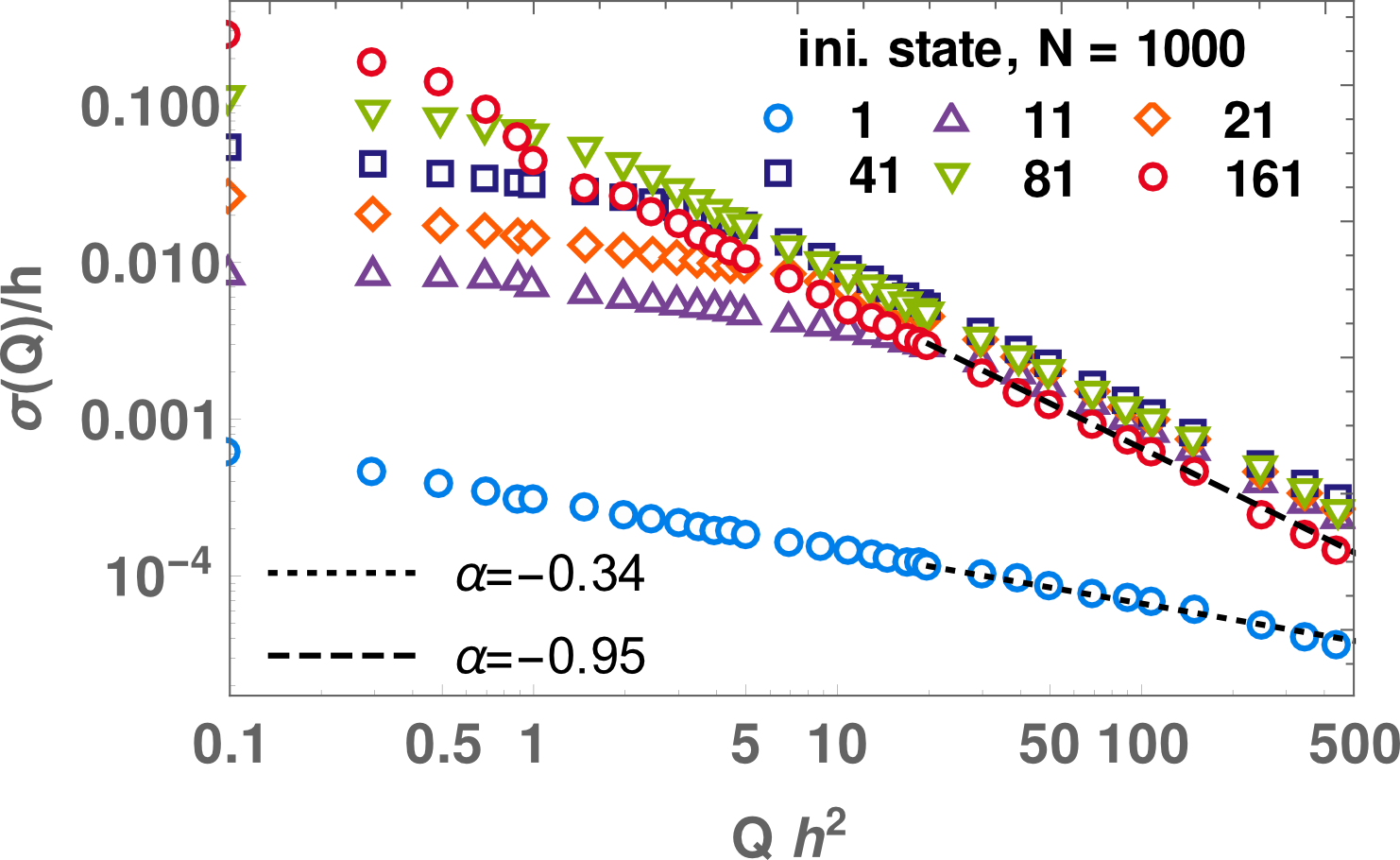}
\includegraphics[width=0.99 \columnwidth]{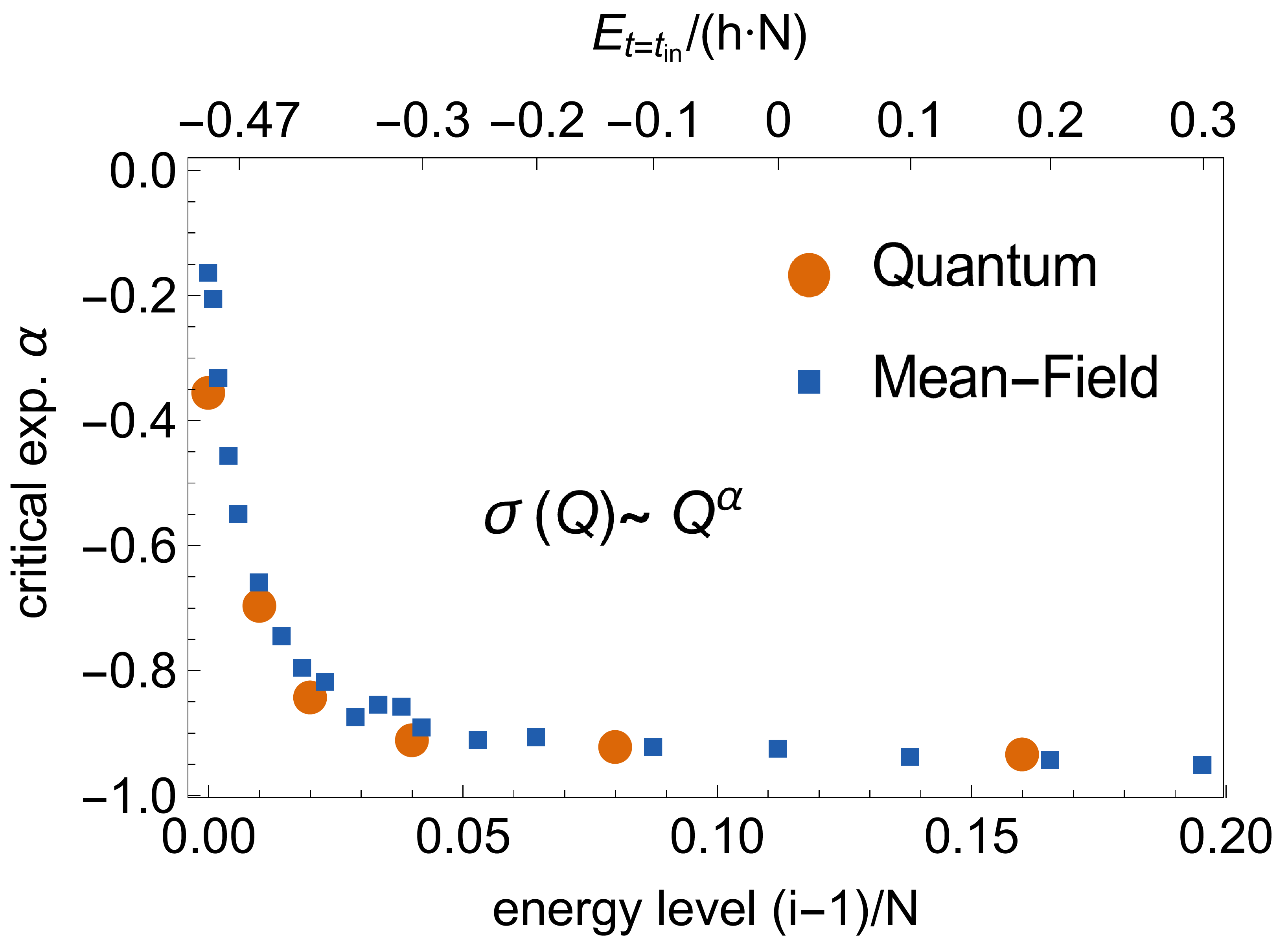}
\caption{(color online)(top panel) The energy distribution width $\sigma(T)$ for different initial eigenstates $\ket{i(t_{\rm in})}$ as a function of the quench speed $Q$ in a double-logarithmic representation. Time $t$ and coupling $\gamma_x(t)$ chosen such that $\gamma_x(t) \approx \gamma_x^{cr,i}$, the critical coupling for crossing the ESQPT line $E/N = -0.5h$ in Fig. \ref{fig:energy_spectrum_gap}. (lower panel) $\alpha$ exponent of the $\sigma(Q)$ curve in the linear area for $Q h^2 >5$ (filled circles). Blue squares represent the corresponding exponents obtained from a mean-field solution, see Sec. \ref{sec:meanfield}. The double top connects the initial state $\ket{i}$ with its energy $E_{t=t_{\rm in}}$. Parameters: $N = 1000$. 
}
\label{fig:Lmg_addiab_scalling}
\end{figure}

For a deeper analysis we now investigate the scaling  of the energy spread $\sigma(t)$ by plotting it for different $Q$ values, keeping  $\gamma_x(t)$ fixed, see Fig. \ref{fig:Lmg_addiab_scalling} (top panel).  For each initial state $\ket{i(t_{\rm in})}$ we choose $\gamma_x(t)$ to be close to their specific transition at $\gamma_x^{cr,i}$, see Fig. \ref{fig:energy_spectrum_gap}. Qualitatively, the scaling of $\sigma(Q)$ is similar for both excited  and ground states, as there are two regions for $Q h^2 <1$ and $Q h^2 \gg 1$ where the scaling is linear in the double logarithmic representation, but with different slopes. A similar separation has been found recently for the Jaynes-Cummings model \cite{Rabi-QPT-addiab_dynam-plenio}.  Quantitatively, the scaling exponent $\alpha$ for $Q h^2 \gg 1$ , which is the slope in the double-logarithmic representation, i.e. 
\begin{equation}
\sigma(Q) \sim Q^\alpha, 
\end{equation}
changes for different initial states, see Fig. \ref{fig:Lmg_addiab_scalling} (lower panel, orange dots). For the ground state $i = 1$ the exponent is around -0.35 and agrees with Ref. \cite{Rabi-QPT-addiab_dynam-plenio} and \cite{Kibble-Zureck-crossover_from_classic_to_quantum-Montanegero} in the thermodynamic limit. Note, that the Ref. \cite{Kibble-Zureck-crossover_from_classic_to_quantum-Montanegero} predicts the '-1/3' scaling in the classical regime, which applies for the LMG model with the well known mean-field, see Sec. \ref{sec:meanfield}. For higher excited states the slope decreases and tends to a constant value around -1. Interestingly, the slope variation can be achieved even for the ground state in quantum critical systems. Thus by varying  the effective system size it can be decreased till the value of -2, which is typical of an adiabatic evolution in presence of a second order QPT \cite{Rabi-QPT-addiab_dynam-plenio,LMG-adiabatic_dynamics_Caneva}. In our case the system size $N$ is fixed and the change is caused by the excited states dynamics which suggests a correspondence between the dynamics around the ESQPT for higher states and the finite size dependent dynamics around the ground state QPT. 
Note, the form of the $\sigma(Q)$ curve echoes the gap scaling of the excited states, see Fig. \ref{fig:energy_spectrum_gap}. The constant exponent for $i/N > 0.05$ again shows the dynamical indistinguishably of the excited states close to the ESQPT.

\section{State dynamics}
\label{sec:3-state-dynamics}
\begin{figure}
\includegraphics[width=1 \columnwidth]{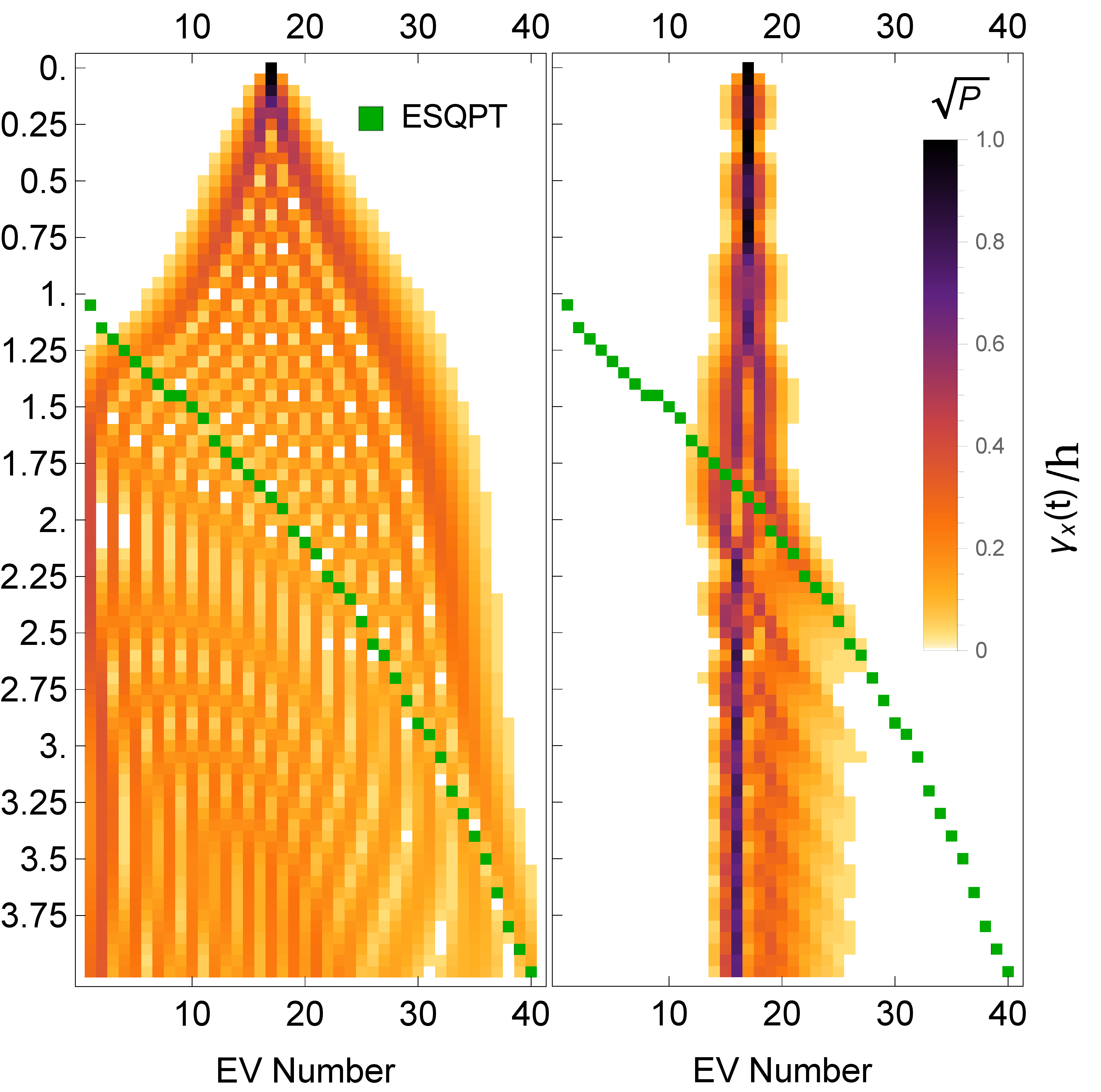}
\caption{(color online) Square root of occupation $\sqrt{P_j}$ of the time-local eigenstates $\ket{j(t)}$ with the number $j$ as a function of time $t$ for the protocol Eq. \eqref{eq:gamma-timedep} with $Q h^2 = 1 $ (left) and $Q h^2 = 10$ (right). The color denotes the contribution of the time-local state to the whole state $\Psi(t)$, thus $P_j(t) = \abs{\alpha_j(t)}^2$. The green dotted line shows where the ESQPT takes place. Note that $\sqrt{P_j}$ for $j>40$ is not shown, as it is zero in both cases. The adiabatic evolution for $\gamma_x(t)<0$ is not shown, too.  Parameters: $N = 200$, initial state at $t_{\rm in}  = 0$ is $\ket{i(t_{\rm in})}=\ket{17}$. }
\label{fig:Lmg_state_occupation_protocol}
\end{figure}

For a further analysis of our protocol, Eq. \eqref{eq:gamma-timedep}, we plot the system dynamics exemplary for two different $Q$ values in Fig.~\ref{fig:Lmg_state_occupation_protocol} by plotting the occupation probability $P_j(t) = \abs{\alpha_j(t)}^2$ for each time-local eigenstate $\ket{j(t)}$. For a small $Q h^2 = 1$ value (left), the adiabaticity is lost shortly after $\gamma_x(t)/h=0$ and the distribution becomes spread between all eigenstates of $H_+(t)$, whereas for a higher $Q h^2=10$ value (right) the adiabaticity is lost at a much later stage $\gamma_x(t)/h \approx 1.25$,  and the wave function is located mostly between the 15.th and 25.th eigenstate. We also recognize, that the width increases continuously for some time after the position of the ESQPT (green dotted line) is crossed \cite{LMG-dynamical_properties_accros_qpt-Mosseri}. This representation suggests, that the ESQPT signatures are visible for higher $Q$ values and become smoothed for the smaller ones, as the wave package becomes then strong delocalized over many eigenstates.

Next, we investigate the dynamics in terms of the spin-coherent states which can be visualized on the Bloch sphere. An instantaneous spin coherent state $\ket{\theta,\varphi}_t \equiv \ket{\theta(t),\varphi(t)}$ rotates the Dicke state $\ket{m=N/2}$ by the angles $\theta$ and $\phi$ and is defined as \cite{Coherent_states-Wigner_distrib_of_a_angular_state-Dowling,Coherent_states-Theory-Zhang}  

\begin{align}
\label{eq:spin-coh-state_def}
\ket{\theta,\varphi}_t &= \rbb{1 + \abs{\exp(i\phi) \tan(\theta/2)}}^{-N/2}   \\
	& \quad  \times \exp\rbb{\exp(i\phi) \tan(\theta/2) \hat{J}_-}\ket{N/2} \notag
\end{align}
and can be rewritten using all eigenstates of the $\hat{J}_z$ as~ \cite{LMG-TC-periodic_dynamic_and_QPT-Georg}
\begin{align}
\ket{\theta,\varphi}_t	&=  \sqrt{\frac{N+1}{4 \pi (\frac{N}{2}+1)}} \notag \\
 & \quad \quad \times \quad \sum_{m=-N/2}^{N/2} 
 \sqrt{\binom{N}{m+N/2}} \cdot \sin\rb{\frac{\theta}{2}}^{\frac{N}{2}+m} \notag \\ 
 &\quad \quad  \times \quad \cos\rb{\frac{\theta}{2}}^{\frac{N}{2}-m} \cdot 
   \exp \rbb{-i \rb{\frac{N}{2} + m}\cdot \varphi} \ket{m}. 
\end{align} 
Thus, the wave function can be expanded in this new basis according to
\begin{equation}
\label{eq:coh_state_wavefunk}
\ket{\Psi(t)} = \int d\theta d\varphi \beta(\theta,\varphi) \ket{\theta,\varphi}_t, 
\end{equation}
where 

\begin{equation}
\beta(\theta,\varphi) = \braket{\Psi(t)}{\theta(t),\varphi(t)}. 
\end{equation}
\begin{figure*}
\includegraphics[width=0.32 \columnwidth]{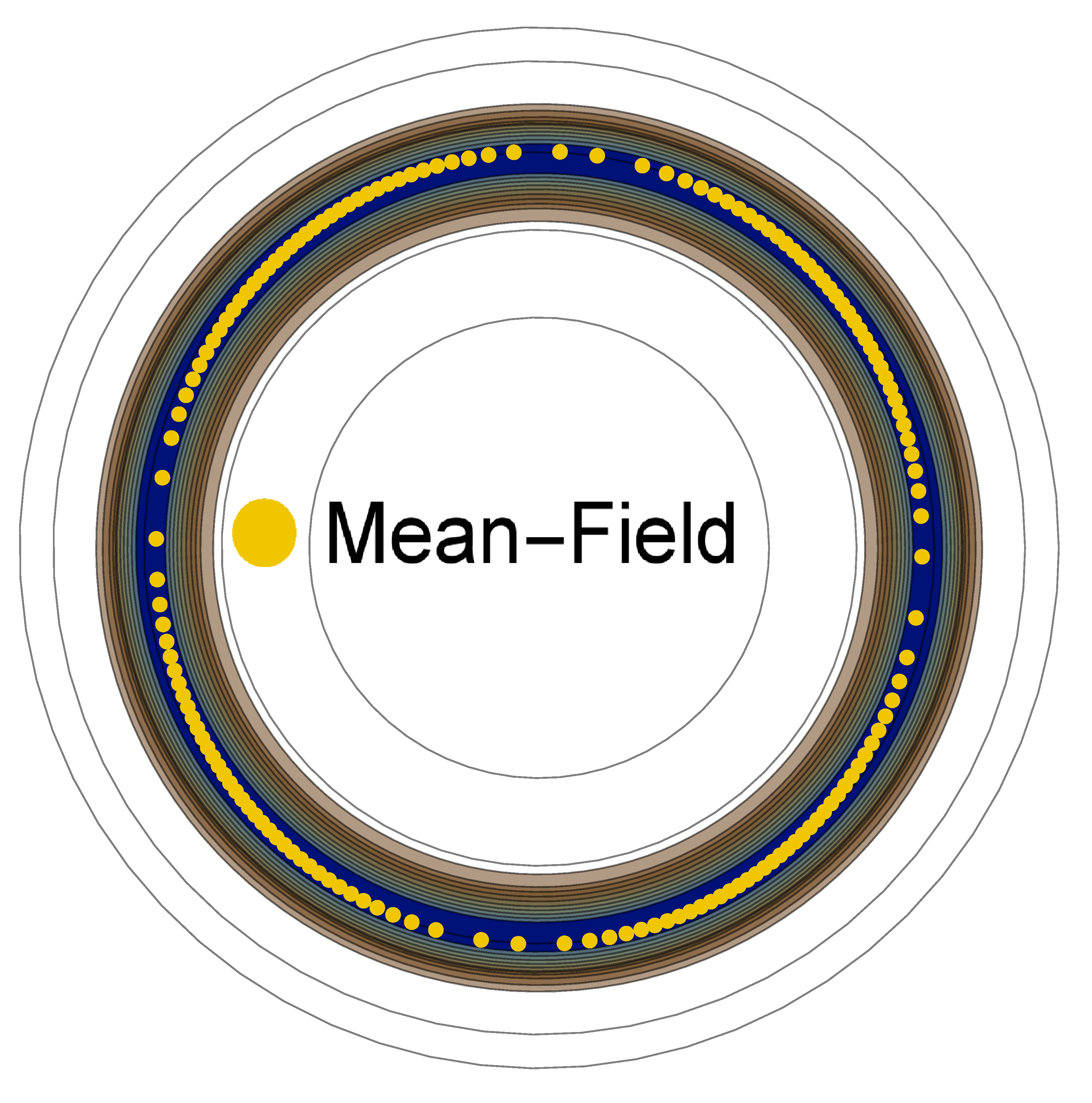}
\includegraphics[width=0.32 \columnwidth]{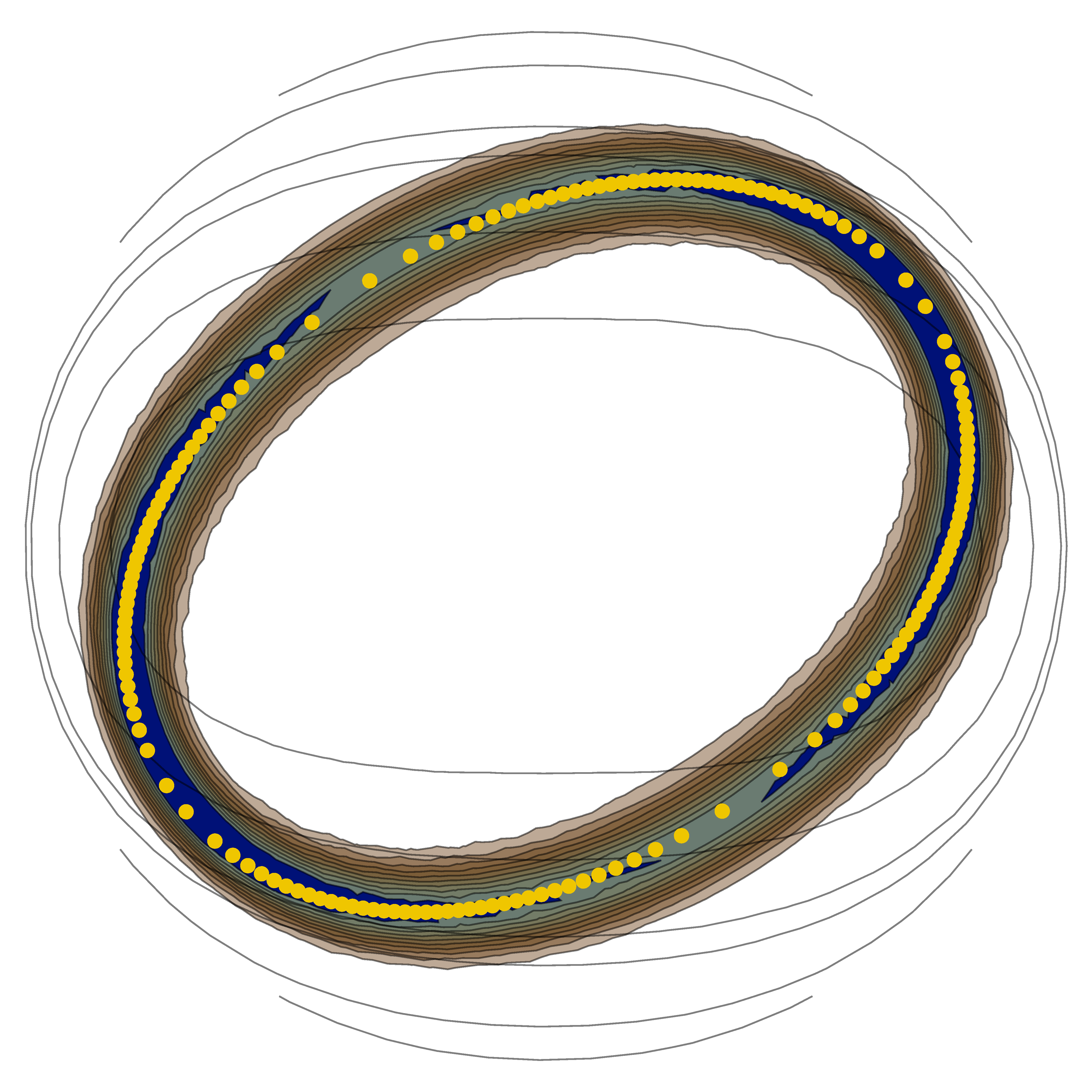}
\includegraphics[width=0.32 \columnwidth]{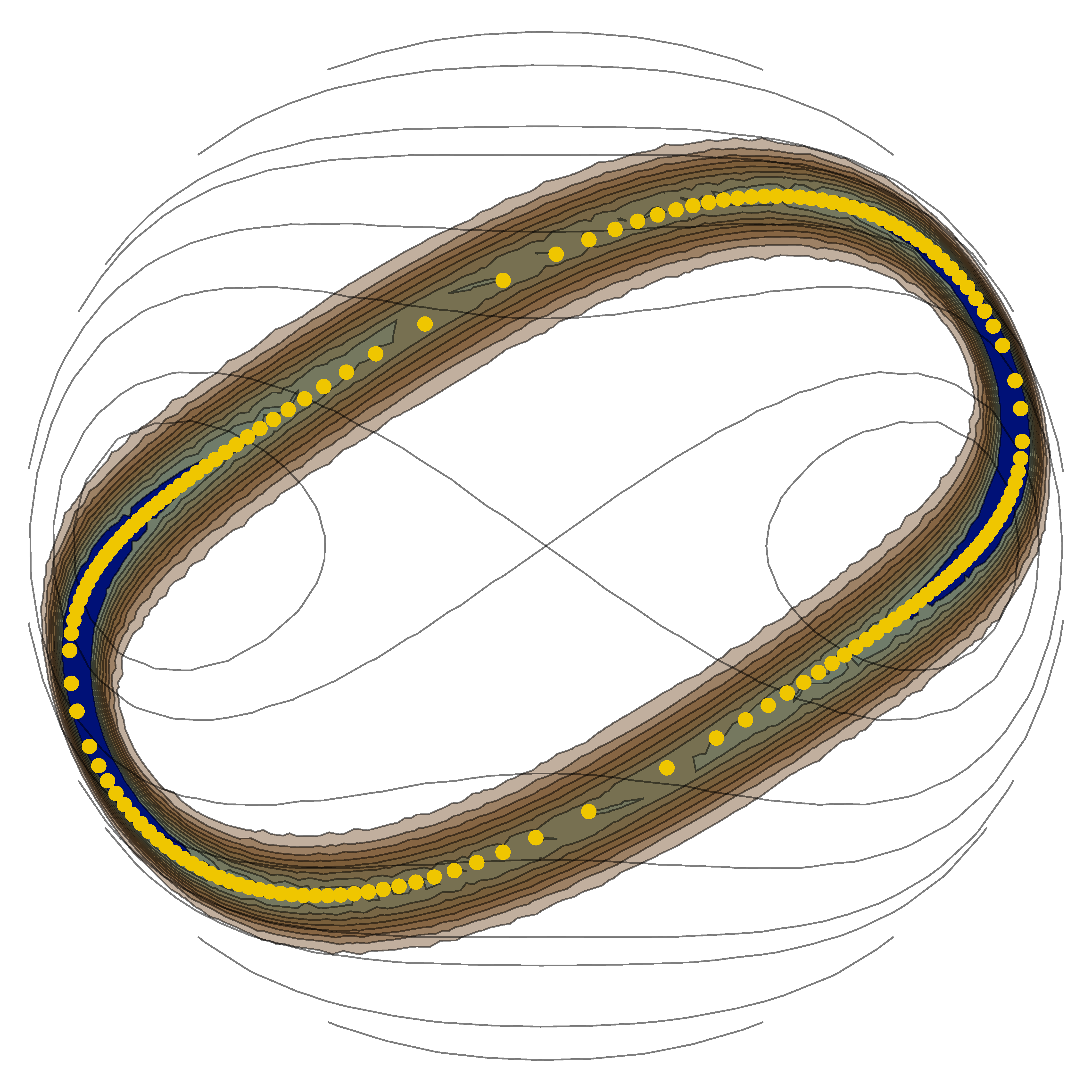}
\includegraphics[width=0.32 \columnwidth]{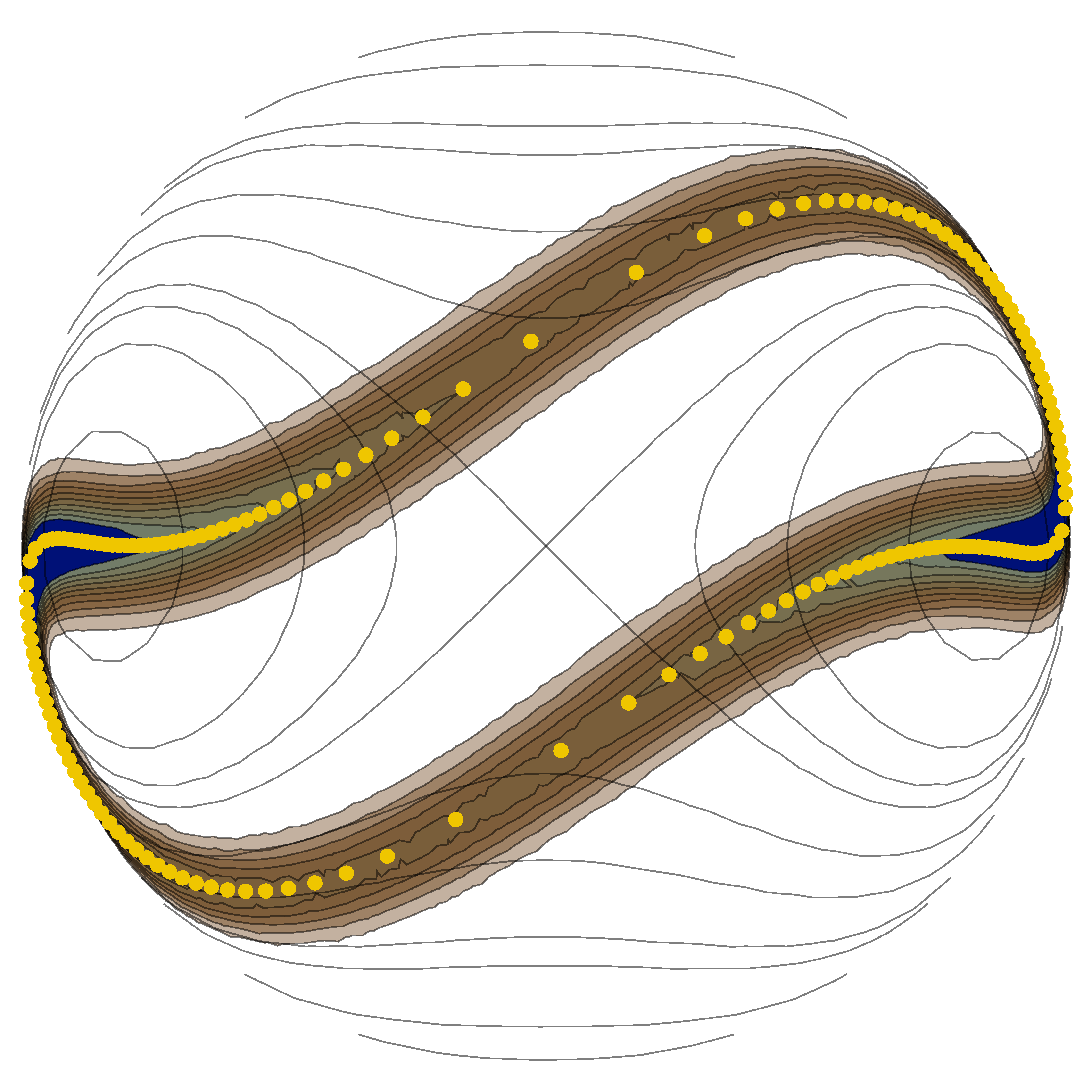}
\includegraphics[width=0.32 \columnwidth]{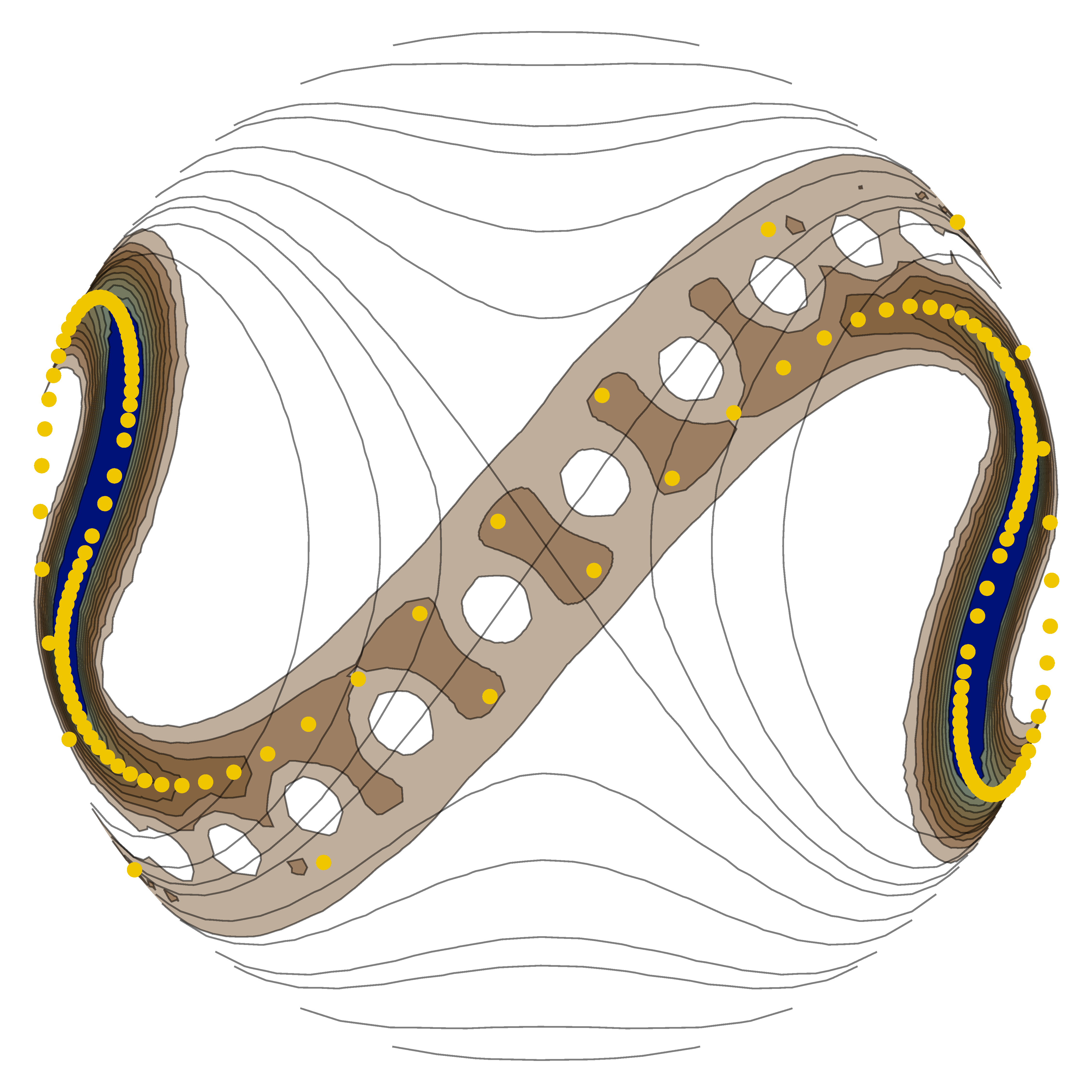}
\includegraphics[width=0.32 \columnwidth]{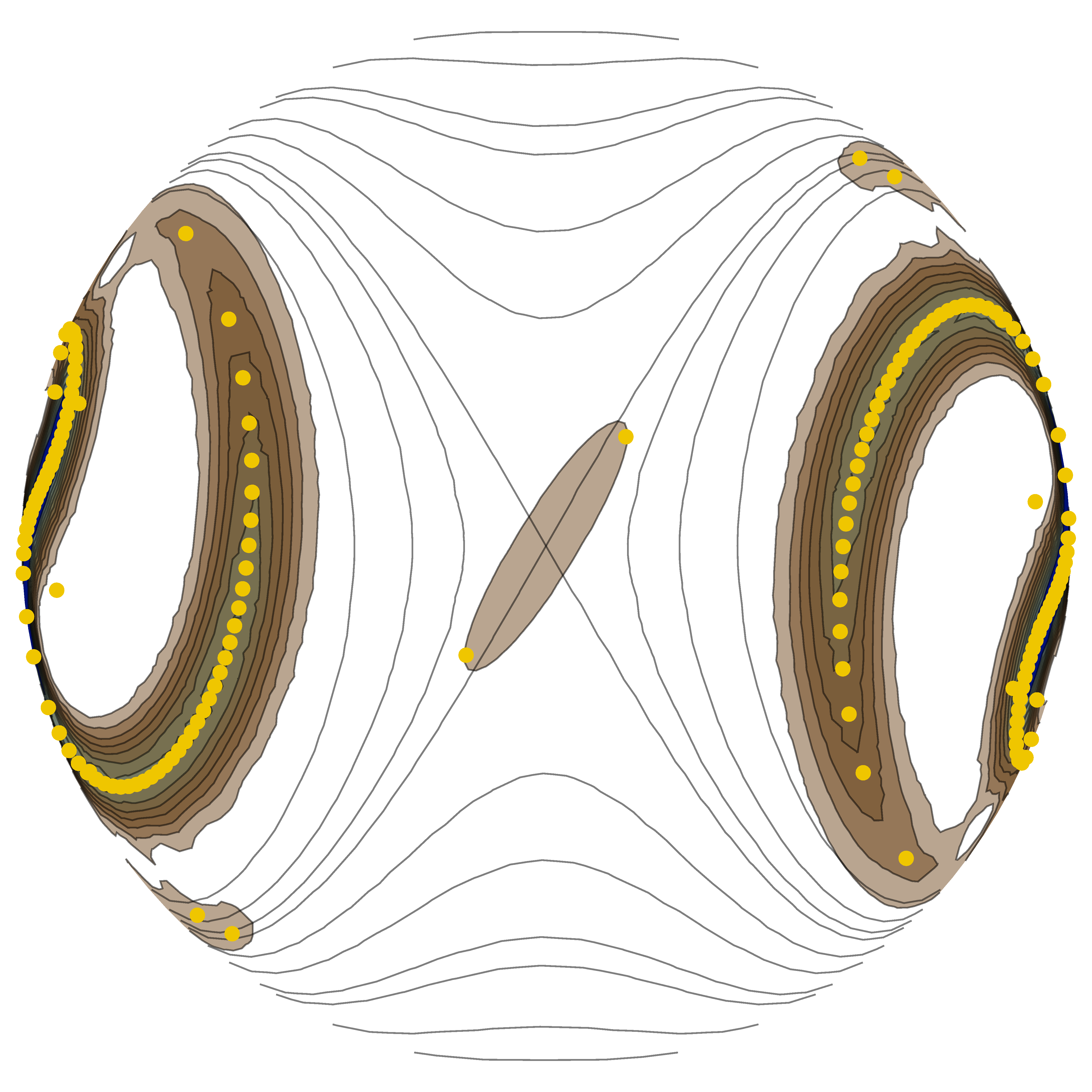}
\includegraphics[width=0.32 \columnwidth]{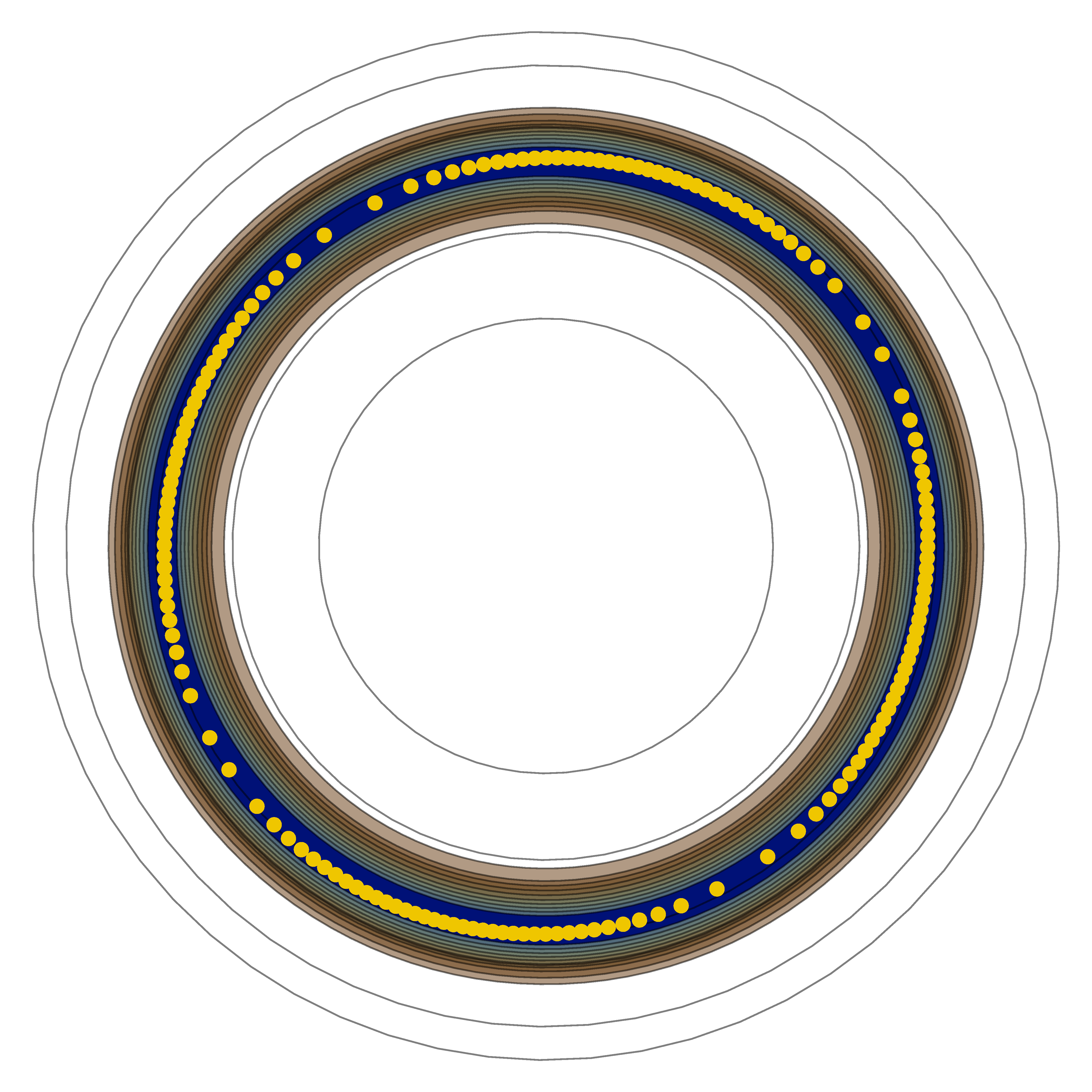}
\includegraphics[width=0.32 \columnwidth]{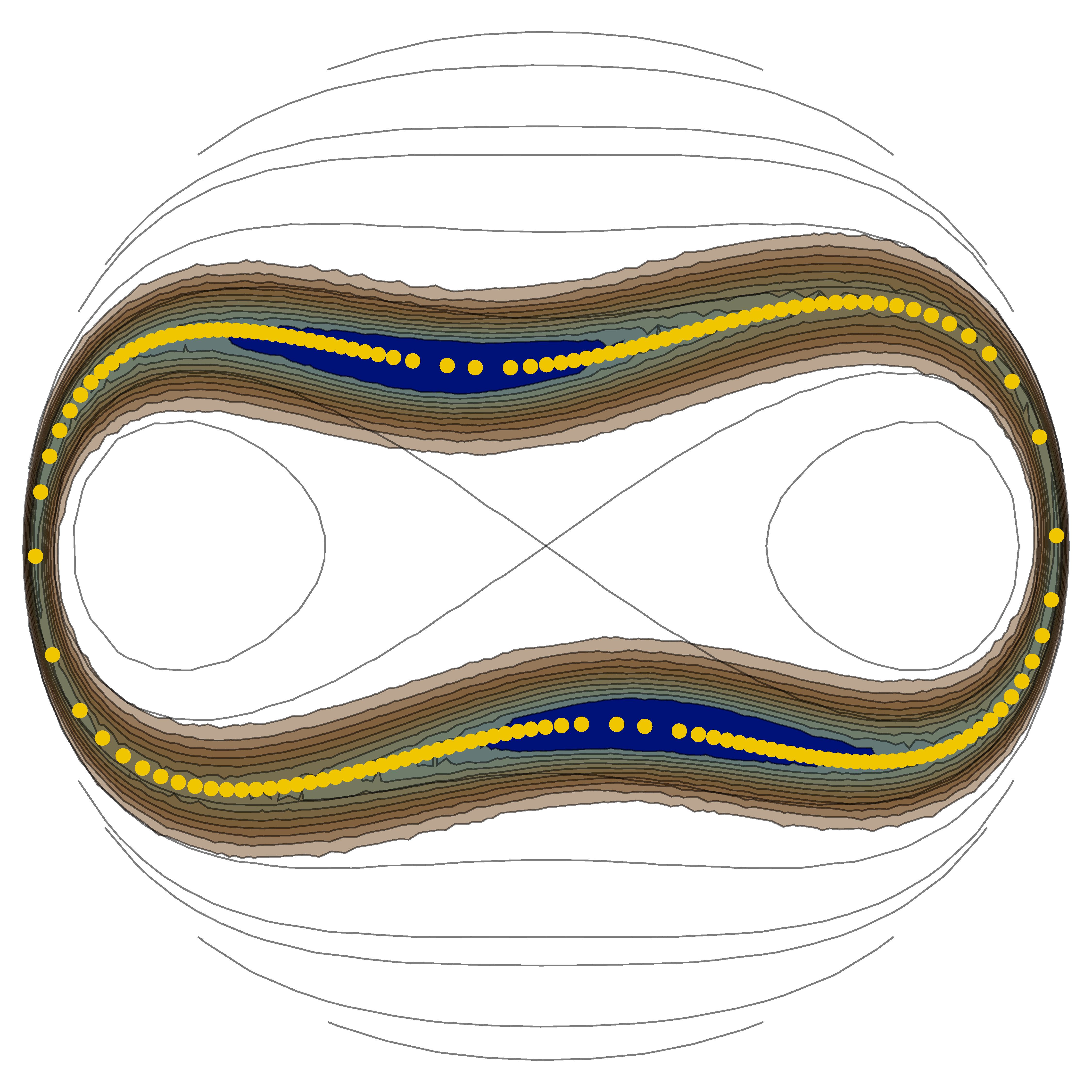}
\includegraphics[width=0.32 \columnwidth]{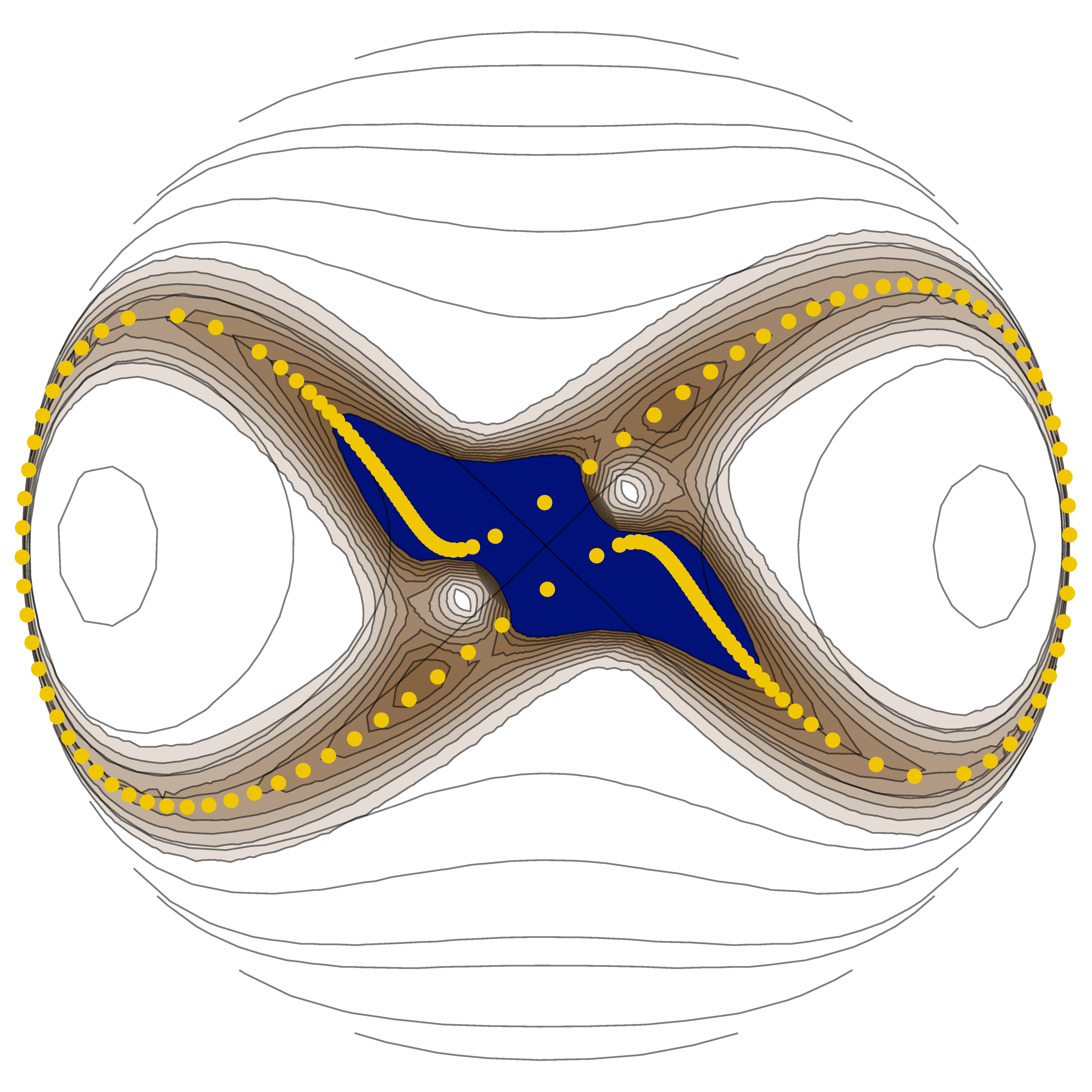}
\includegraphics[width=0.32 \columnwidth]{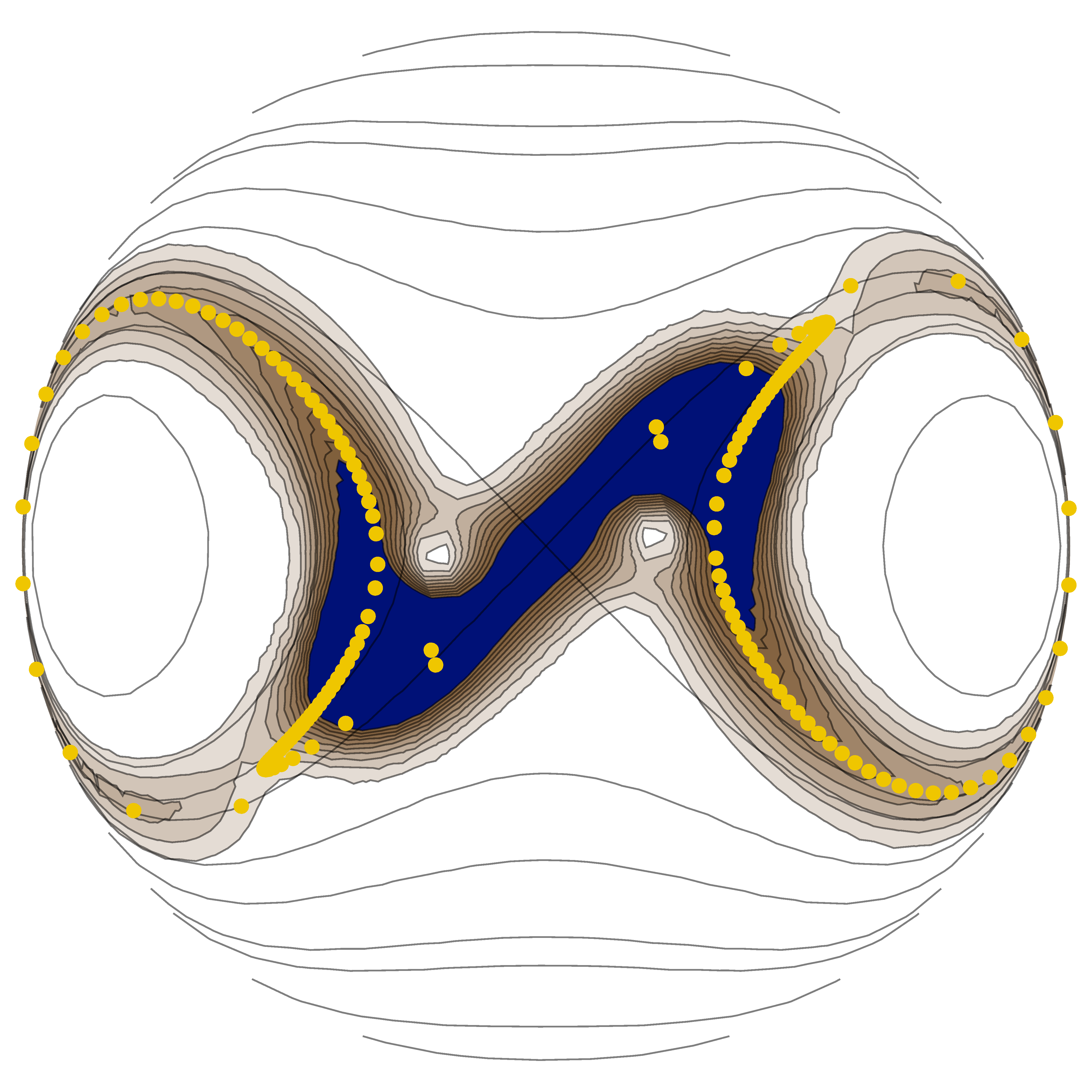}
\includegraphics[width=0.32 \columnwidth]{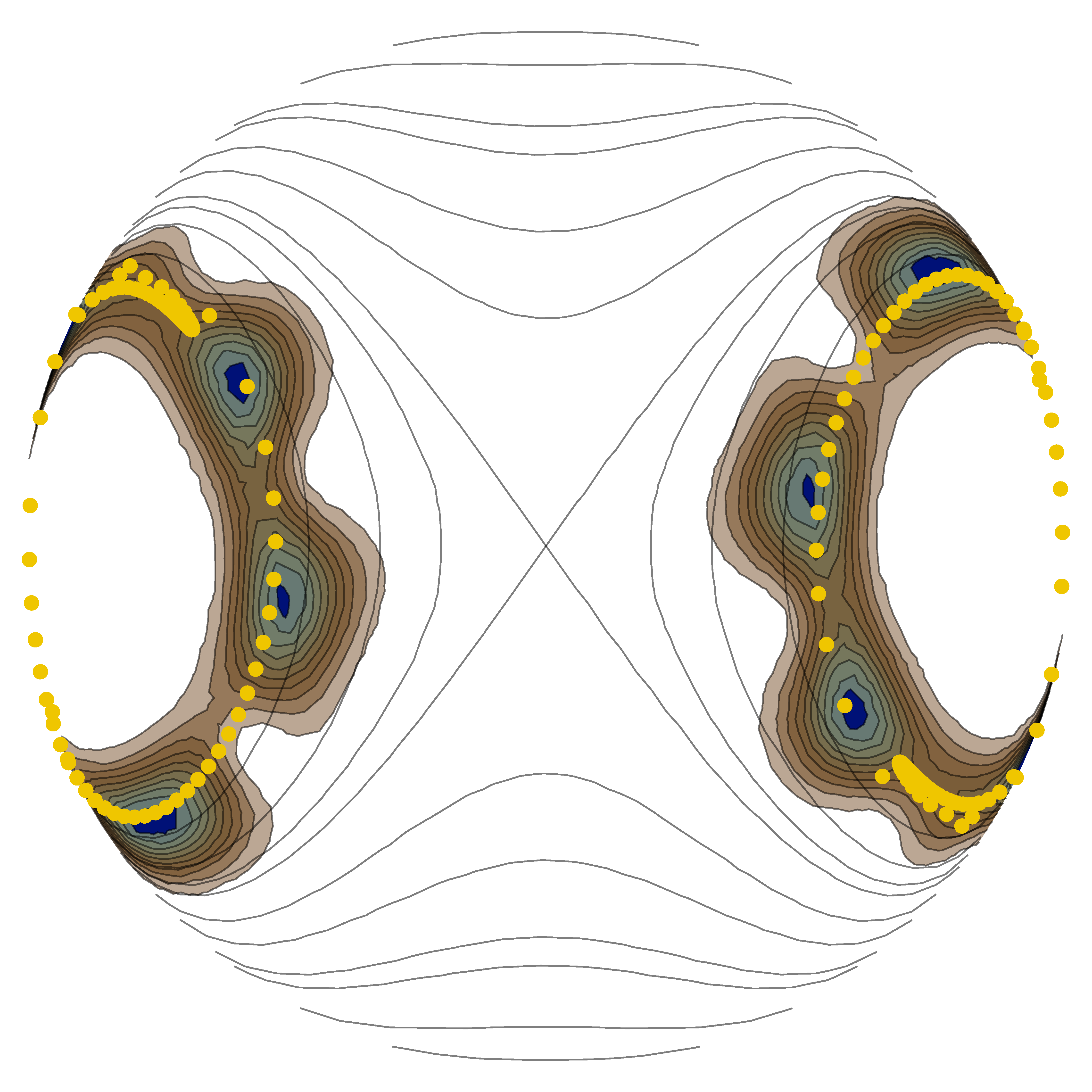}
\includegraphics[width=0.32 \columnwidth]{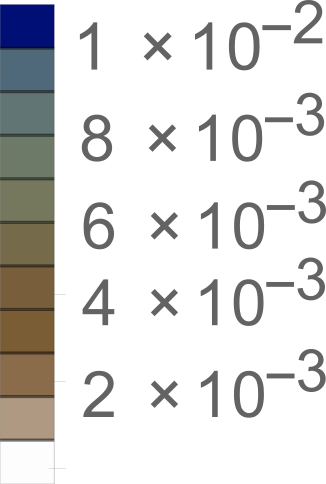}
\caption{(color online) Protocol-dependent development of the wave function in terms of coherent states $\ket{\phi,\theta}$. The results are shown on the upper half of the Bloch sphere  projected to its great circle. The color shows the occupation probability of the correspondent coherent state. Gray lines shows the states with the same energy. The yellow dots show the mean-field solution of corresponding semi-classical equations, see \ref{sec:meanfield}.  Each point corresponds to one coherent state. Parameters: $N = 200$, (upper) $Q h^2=1$, from left to right $\gamma_x(t)/h$ has following values: 0, 1, 1.5, 2, 3, 4. (lower panel) $Qh^2=10$, from left to right $\gamma_x(t)/h$ has following values: 0, 1.5, 1.93, 2.1, 3.}
\label{fig:LMG_blochsphere_dynamics}
\end{figure*}
The quantity $\abs{\beta(\theta,\varphi)}^2$ gives the contribution of the corresponding coherent state to the wave function $\ket{\psi(t)}$. Figure \ref{fig:LMG_blochsphere_dynamics} shows snapshots of this probability for discrete $t \sim \gamma_x(t)$ values on the upper half of the Bloch sphere which is projected to its great circle. The gray lines show the position of coherent states with the same energy. Each of those lines, if the eigenstates were presented directly on the Bloch sphere, would mark the position of the center of a time-local eigenstate with the corresponding energy of the Hamiltonian $\hat{H}$ at time $t$. Compare therefore yellow dots and colored halo in the top left panel of Fig. \ref{fig:LMG_blochsphere_dynamics}. 

The QPT of the ground state is visible in the structure change of the iso-energetic lines, thus for $\gamma_x(t)/h>1$ two different energetic structures are present which are separated by a separatrix shaped as the symbol '$\infty$'. At the separatrix the ESQPT takes place. The separatrix grows with increasing $\gamma_x(t)$. The upper and lower figure panels show the evolution of the wave function in terms of spin coherent states for two different $Q$ parameters (the same as in Fig. \ref{fig:Lmg_state_occupation_protocol}) , one smaller (top panel, fast quench) and larger (lower panel, slow quench). The initial state is in both cases the same as in Fig. \ref{fig:Lmg_state_occupation_protocol}, thus the 17th excited eigenstate of the Hamiltonian $\hat{H}_+(\gamma_x(t=0))$ and coincides with one gray line in Fig. \ref{fig:LMG_blochsphere_dynamics}. Note that the width of the $\beta(\theta,\varphi)$ distribution becomes smaller with larger $N$. The adiabaticity loss due to the appearance of the QPT and the ESQPT at some later point in the dynamical evolution is visible by a deviation of the coherent state distribution from the corresponding gray line. For a faster protocol, the difference is clearly visible at $\gamma_x(t)/h=1$ (second figure, top panel), whereas for a slower protocol the wave package approximately follows the corresponding gray energy line even at $\gamma_x(t)/h = 1.5$ (second figure, lower part). However, the protocol speed is essential for a further evolution.  For a fast evolution (top panel) the wave package does not have enough time to adapt and keeps its shape till $\gamma_x(t)/h \approx 2$ (first four snapshots). Meanwhile the separatrix goes through the wave package and embeds it for higher $\gamma_x(t)$ values. For a slower change of $\gamma_x(t)$ (lower part) the wave package approximately still follows the corresponding energy line which collects it around the north pole on the Bloch sphere along the separatrix line (see third and fourth snapshot). Due to this fact the ESQPT signatures should be present for example in observables like $\avg{J_z}(t)$ or $\avg{J_x^2}(t)$. For a higher $\gamma_x(t)$ value the wave package is completely embedded by the separatrix.

\section{Mean-field connection}
\label{sec:meanfield}
The coherent state representation is closely connected to the mean-field treatment, which in case of the closed LMG model coincides with the quantum results in the thermodynamic limit \cite{LMG-spectrum_thermodynamic_limit_and_finite_size-corr-Mosseri}. In this section we apply the protocol Eq. \eqref{eq:gamma-timedep} at the mean-field level and compare the resulting semi-classical with the numerical quantum results from the previous sections. However, the mean-field provides the system information based on the operator expectation values and it is not possible to fix some energy level just by counting from the ground state, which we have explicitly used in the previous section.    

To derive the mean-field equations we use the Heisenberg equation $\partial_t \avg{\hat{J}_\eta} = i\avg{\commut{\hat{H},\hat{J}_\eta}}$ and assume that the factorization $\avg{\hat{J}\eta \hat{J}_\eta'}=\avg{\hat{J}_\eta}\avg{\hat{J}_\eta'}$ for $\eta \in {x,y,z}$ holds in the thermodynamic limit, this yields  \cite{Morrison-Dissipative_LMG-and_QPT,LMG-Finite_size_scalling_Dusuel}

\begin{align}
\label{eq:lmg_mean_field}
\dot{J}_x(t) &= h J_y(t), \\
\dot{J}_y(t) &= -h J_x(t) + 2 \gamma_x(t) J_x(t) J_z(t),\\
\dot{J}_z(t) &= -2 \gamma_x(t) J_x(t) J_y(t),
\end{align}

where $J_i(t) \equiv \frac{1}{N}\avg{\hat{J}_i}$ denotes the rescaled average operator value.

To compare the results between the fully quantum and the mean-field calculations, we have to start with a similar initial condition in both cases. However, the initial condition in case of the mean-field calculation is a point on the Bloch sphere and corresponds to a spin-coherent state. In contrast, in the quantum case we have chosen an eigenstate of the Hamiltonian $\hat{H}_+$ at $t=t_{\rm in}$ as the initial condition. But an eigenstate is according to Eq. \ref{eq:coh_state_wavefunk} a superposition of spin coherent states, see Fig. \ref{fig:LMG_blochsphere_dynamics} for the visualization. Therefore, in the mean-field case we start with a set of initial states, whose energies correspond to the eigenenergy of the initial eigenstate in the quantum case we want to compare with. The dynamical evolution of coherent states is shown in Fig. \ref{fig:LMG_blochsphere_dynamics} by yellow dots. We see that the quantum dynamics of the wave functions is captured by the mean-field calculation very well. 

In case of the adiabatic evolution the energy of each coherent state in the set would evolve in the same way, thus the width of the energy distribution for a set of coherent states would be zero. However, in case of the adiabaticity breakdown the set of coherent states will evolve in a different manner and as a consequence the width of the energy distribution becomes finite. We will characterize the width by the standard deviation $\sigma_M$ and calculate it as a function of $Q$ for sets of coherent states with different initial energy $E_{\rm in} \equiv E(t = t_{\rm in})$ at a special time $\gamma_x^{cr,E_{\rm in}}(t)$. The latter is equal to $\gamma_x^{cr,i}$ if $\bra{i(t_{\rm in})}\hat{H}_+\ket{i(t_{\rm in})}=E_{\rm in}$ and gives the value  when in case of the fully adiabatic evolution the initial set will cross the ESQPT region. Same as in the quantum case, the curve $\sigma_M(Q)$ has a linear part for $Q h^2 \gg 1$ in the double-logarithmic representation (not shown here), which we fit by a power law 
\begin{equation}
\sigma_M(Q) \sim Q^\alpha.
\end{equation}
The corresponding exponents $\alpha$  for different initial energies are shown in Fig. \ref{fig:Lmg_addiab_scalling} by blue squares and matches pretty well to the purely quantum calculations. However the perfect agreement can not be expected due to the finite-size effects on the one hand and the factorization assumption at the lowest level on the other.

\begin{figure}
\includegraphics[width=1 \columnwidth]{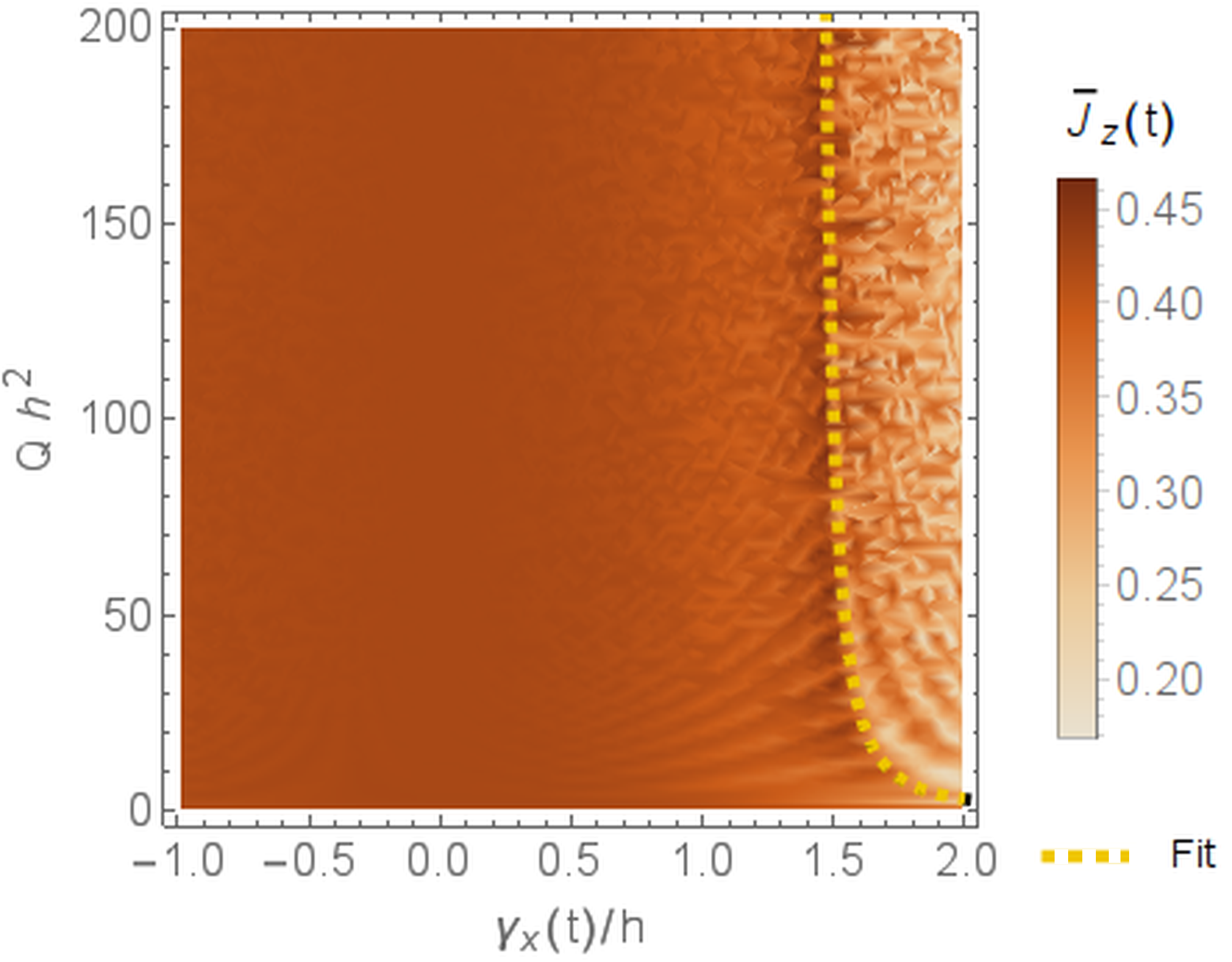}
\caption{(color online) The $\bar{J}_z$ expectation value based on the mean-field theory calculated by averaging over a set of coherent states with a fixed energy $E_{\rm in}/(h \cdot N)$ at the initial time $t=t_{\rm in}$ in the $(Q h^2,\gamma_x(t))$ plane. The increase of $J_z$ value is connected to the passing of the ESQPT region in the system, see blue line. Parameters: $E_{\rm in}/(h \cdot N) = -0.265$, $t_{\rm in} = -1/h$. }
\label{fig:LMG_jz}
\end{figure}

The ongoing protocol deforms the energy landscape (see Fig. \ref{fig:LMG_blochsphere_dynamics}), induces a symmetry breaking and creates a separatrix for $\gamma_x(t)/h >  1$.  Especially for a slower protocol (large $Q$) the set of initial states has a small 'energy width' and goes around the same time through the growing separatrix. However, this corresponds to a system which goes through the ESQPT. This is visible in the observables behavior as a typical peak, in case of $J_z$ the peak arises around the value of '0.5'. In Fig. \ref{fig:LMG_jz} we show the $\bar{J}_z$ expectation value averaged over the set of different initial conditions with the same energy $E_{\rm in}$, thus over the $J_z$ values of yellow dots in Fig. \ref{fig:LMG_blochsphere_dynamics}, for different protocol speeds $Q$ as a function of $\gamma_x(t)$. Such averaging helps to suppress protocol-induced oscillations. 
One can see clearly a sharp increase (yellow dotted line) of $\bar{J}_z$ which occurs when the wave package comes close to the separatrix. For lower $Q$ the adiabaticity breaks down and the peak is moved to higher $\gamma_x$ values  becoming more flat. For higher $Q$ values the position of the peak tends to $\gamma_x(t)/h = 1.4$, which corresponds to the ESQPT position for the given initial energy $E_{\rm in}$.  We could fit the $Q$ dependent position of the peak by $Q_{peak} = 1/(\gamma_x(t)-1.4)^{2}$, where the exponent '-2' was the fitting parameter. Note that we have also checked that the quantum calculation leads to the quantitatively same results.



\section{Summary}
\label{sec:5-summary}
In this work we have investigated the role of the ESQPT for the adiabaticity loss of the excited states dynamics in the LMG model. Similar to the adiabaticity loss of the ground state evolution close to the QPT, the energy spectrum around the ESQPT in case of the LMG model is composed of a series of avoided crossings, which will make the dynamics of excited states non-adiabatic. To probe the adiabaticity loss, we have linearly increased the internal coupling in the model and crossed the ESQPT boundary by starting in different excited states. We argue, that the usually used residual energy to quantify the adiabaticity loss is not useful for the excited states, therefore we investigate the energy width of the wave package.  Our results show that even for the excited states the energy width of the wave package close to the ESQPT scales with the power law as a function of the inverse protocol speed $Q$. Though, the corresponding exponent depends on the excited state. Whereas for the lower part of the spectrum there is a significant change in the exponents, for the higher part of the energy spectrum this exponent tends to a constant value. 

Additionally, we showed that the mean-field and quantum based results agree. Therefore, we linked the eigenstates of the LMG Hamiltonian with a set of coherent states, which are the closest ones to the mean-field states.  Thus, we investigated the evolution of the mean-field states set with initially same energy. Due to their different position on the Bloch sphere, each of those states evolves in a different manner under a protocol action. The deviation of the fully adiabatic evolution of this mean-field dynamics gives on average the same scaling behavior as a function of the protocol speed. In the Bloch sphere representation the protocol action can be understood as passing of the protocol-caused separatrix  growth cutting through the wave package. Especially for a slower protocol evolution the ESQPT signatures can be identified in the dynamical evolution of the wave package.

\section*{Acknowledgments}\vspace{-2mm}
We thank Georg Engelhardt for useful discussions. The authors gratefully acknowledge financial support from the DFG Grants BR 1528/9-1, SFB 910, GRK 1558.

\end{document}